\documentclass[a4paper,11pt]{article}
\usepackage{jheppub} 
\usepackage{orcidlink}

\usepackage{booktabs}
\usepackage{tabularx}

\usepackage{graphicx}
\usepackage{dcolumn}
\usepackage{comment}
\usepackage{bm}
\usepackage[mathlines]{lineno}

\newcommand{\df}[1]{{\color{red}{[{\bf DF:} #1]}}}

\long\def\exclude#1{}

\title{The imprint of neutron escape on non-thermal protons in AGN coronae}

\author[a,b]{Antonio Ambrosone
\orcidlink{0000-0003-0205-3010},}
\affiliation[a]{Gran Sasso Science Institute (GSSI), Viale Francesco Crispi 7, 67100 L’Aquila, Italy}
\affiliation[b]{INFN-Laboratori Nazionali del Gran Sasso (LNGS), via G. Acitelli 22, 67100 Assergi (AQ), Italy}

\author[a,c]{Damiano F.\ G.\ Fiorillo \orcidlink{0000-0003-4927-9850},} 
\affiliation[a,c]{Istituto Nazionale di Fisica Nucleare (INFN), Sezione di Napoli,
Complesso Universitario di Monte Sant’Angelo, Via Cintia, 80126 Napoli, Italy}

\emailAdd{antonio.ambrosone@gssi.it}
\emailAdd{damiano.fiorillo@gssi.it}

\abstract{
The coronae of active galactic nuclei (AGN) are promising sites of proton acceleration to hundreds of TeV, potentially accounting for the neutrino excess from NGC 1068 reported by IceCube. We show that photohadronic proton-to-neutron conversion provides an important proton escape channel, as neutrons can leave the corona well before decaying. In magnetized turbulent coronae, neutron escape can reduce the characteristic neutrino energy by up to a factor of four  and the integrated neutrino luminosity by approximately \(30\%\) at a fixed non-thermal proton energy density. 
More strikingly, proton-to-neutron conversion offers an escape route roughly two orders of magnitude faster than proton diffusion, requiring a correspondingly larger proton injection rate to sustain the non-thermal population. The effect is considerably weaker when proton escape is already ballistic, as expected in magnetic-reconnection scenarios. Therefore, neutron escape is  essential for quantitatively connecting neutrino emission from AGN coronae to the efficiency of proton injection into the acceleration process.}

\begin{document}

\maketitle
\flushbottom


\section{Introduction}~\label{Sec:intro}

A diffuse extragalactic neutrino flux, with energies ranging from a few~TeV to several~PeV, was discovered by the IceCube Collaboration in 2013, opening a new window on the non-thermal Universe~\cite{IceCube:2013low}. Unlike photons and charged cosmic rays~(CRs), neutrinos propagate over cosmological distances without absorption or deflection, carrying direct information about their production sites. Neutrinos escape freely even from those cosmic accelerators where gamma radiation is strongly attenuated, often termed hidden or gamma-ray opaque sources. 

These hidden sources play a major role because of the mismatch between the large neutrino flux observed at a few TeV, in comparison with the Fermi-LAT measurements~\cite{Fermi-LAT:2014ryh} of the extragalactic diffuse gamma-ray flux at hundreds of GeV. The natural explanation is that the neutrino flux primarily arises in hidden sources~\cite{Murase:2015xka,Chang:2016ljk,Xiao:2016rvd,Sudoh:2018ana,Capanema:2020rjj,Capanema:2020oet,Fang:2022trf} (see also \cite{Ambrosone:2020evo} and references therein). If neutrinos are produced through the interaction of non-thermal protons with a dense radiation field, that same field offers a natural target to attenuate gamma rays~\cite{Murase:2015xka}, realizing the hidden source scenario suggested by multimessenger observations.

An independent hint towards the hidden-source paradigm has come from the IceCube discovery of a neutrino excess directionally coincident with an active galactic nucleus (AGN), the Seyfert galaxy NGC~1068~\cite{IceCube:2022der}. No comparable gamma-ray flux has been observed, with the strongest upper bound by the MAGIC telescope roughly an order of magnitude lower~\cite{MAGIC:2019fvw}, so neutrinos are expected to be produced in a compact, optically thick region where gamma rays can be absorbed~\cite{Inoue:2019yfs,Kheirandish:2021wkm,Murase:2022dog,Eichmann:2022lxh,Mbarek:2023yeq,Fiorillo:2023dts,Karavola:2024uui,Fiorillo:2024akm,Lemoine:2024roa,Ambrosone:2024zrf,Fiorillo:2025ehn,Yuan:2025ctq,Carpio:2026xkf,Testagrossa:2026jcs,Passos-Reis:2026owu}. A natural candidate is the corona surrounding the central black hole of the AGN, exhibiting a prominent X-ray emission up to hundreds of keV which naturally absorbs gamma rays between GeV and TeV energies~\cite{Laha:2024tnc}.

Even after specifying the candidate region for non-thermal emission, the dynamics of non-thermal particles remains largely uncertain. Neutrinos are emitted by the decay of high-energy pions produced in non-thermal proton interactions (see Ref.~\cite{Das:2026rwt} for a critical discussion of alternative scenarios), crucially depending on what mechanism accelerates non-thermal protons, and their dominant interactions. The spectral properties of the neutrino excess observed by IceCube, showing a neutrino spectrum steeply falling after a few TeV, may offer some guidance in constraining candidate scenarios. Currently, several possibilities consistently explain the broad properties of the signal, including diffusive acceleration in a coronal shock~\cite{Inoue:2019yfs}, stochastic acceleration in magnetized turbulence~\cite{Murase:2019vdl,Kheirandish:2021wkm,Fiorillo:2024akm,Testagrossa:2026jcs}, and a magnetized reconnection layer in the black hole magnetosphere~\cite{Fiorillo:2023dts,Karavola:2024uui,Karavola:2026rpg}.

Associated with the uncertain acceleration mechanism, the coronal size also affects the nature of hadronic interactions; a relatively wide corona spanning hundreds of gravitational radii is permeated by the strong optical-ultraviolet (OUV) field produced from the surrounding disk, so Bethe--Heitler~(BH) pair production $p\gamma\to p e^+ e^-$ strongly cools the protons~\cite{Murase:2019vdl}. A more compact corona spanning up to a few tens of gravitational radii is likely less affected by the diluted OUV radiation produced in much outer regions~\cite{Fiorillo:2023dts,LeBihan:2026rqq, Lemoine:2024roa}, so the dominant proton cooling may come from $p\gamma$ interactions with the X-ray coronal field.

In most  of the scenarios discussed above, pions are predominantly produced in $p\gamma$ interactions. The latter thus play two roles: first, they furnish a means of neutrino production through pion decay; second, they drain the non-thermal proton population of energy. The latter feature is quite consequential, since it competes with the escape and acceleration of protons, shaping their spectrum. Usually, $p\gamma$ interactions are treated as a continuous energy loss, e.g.\ in Refs.~\cite{Kimura:2014jba,Murase:2019vdl,Kheirandish:2021wkm,Fiorillo:2024akm,Lemoine:2024roa,Fiorillo:2025ehn,Yuan:2025ctq,Testagrossa:2026jcs}. However, inelastic photohadronic interactions turn protons into neutrons~\cite{Fedynitch:2025mgj,Owen:2026ifv}, which are not magnetically confined in the corona, and in fact escape much faster than they can decay back to protons. Thus, proton-to-neutron conversion via $p\gamma$, beyond cooling protons, offers an effective escape channel to the otherwise confined protons~(see for instance early works by Refs.~\cite{1990ApJ...362...38B, 1989ApJ...341L..33S, 2012ApJ...754..148T} where the impact of neutrons on the AGN dynamics is discussed).

Here we consider the impact of this effect on stochastic acceleration in magnetized coronae. To do this, we consider the coupled evolution of protons and neutrons, rather than considering the cooled protons alone. Our methods are borrowed from the trickle-down~(TD) approach applied in the ultra-high-energy regime~\cite{Unger:2015laa}, where protons and neutrons are catastrophically removed from their population every time they interact through $p\gamma$ scattering, and re-injected at lower energies according to their interaction probabilities.
We consider the impact of proton-to-neutron conversion both in models of stochastic proton acceleration in magnetized turbulence, and magnetized reconnection. The key effects are a distortion of the neutrino spectrum, due to the increased losses suffered by protons in inelastic interactions, and an enhanced proton escape channel. For magnetized turbulence, the latter effect is quite dramatic, since protons, which would otherwise be magnetically confined in the turbulent region for days, can convert to neutrons and escape the region at the speed of light roughly in a few minutes. Consequently, a much larger rate of injected protons is required to saturate the IceCube neutrino signal, although the overall number density of non-thermal protons is still much smaller than the available baryon density. For a magnetized reconnection layer, where protons anyway escape within roughly a light-crossing time, the impact of proton-to-neutron conversion is more limited.

The article is organised as follows. In Sec.~\ref{sec:physical_setup}, we introduce the background photon fields, the coronal geometries, and the relevant cosmic-ray timescales. In Sec.~\ref{sec:cr_evolution},  we discuss the transport equations for both the magnetized turbulent-corona and magnetic-reconnection models for the continuous-energy-loss~(CEL) approximation and the trickle-down approach, respectively. Sec.~\ref{subsec:neutrino_production} discusses how we implement the neutrino production rates after determining the equilibrium densities of cosmic rays. 
 In Secs.~\ref{Sec:results_cosmic_rays_magnetized}
and~\ref{Sec:magnetic_reconnection}, we present the results
for the two models, respectively. In Sec.~\ref{Sec:conclusions}, we discuss the implications of our findings and draw our conclusions.
 Appendices~\ref{sec:app_photopion} and \ref{sec:app_hadronuclear} provide details on the photopion and hadronuclear simulations and the calculation of the interaction quantities employed in this work, respectively. Appendix~\ref{app:numerical_transport} describes the numerical implementation and the boundary conditions used to solve the transport equations.  Finally, in Appendix~\ref{app:secondary_cooling}, we assess the impact of secondary cooling including synchrotron losses of charged pions and muons on the neutrino yields.


\section{Cosmic-ray dynamics in coronal environments}
\label{sec:physical_setup}

The dynamics of protons depends rather strongly on the environment---how strongly magnetized the corona is, and the radiation field that protons experience inside of it. We discuss these model choices in this section, starting from the radiation field and then particle acceleration, escape and energy losses mechanisms.

\subsection{Radiation field }

\begin{figure}[h!]
\centering
\includegraphics[width=0.9\linewidth]
{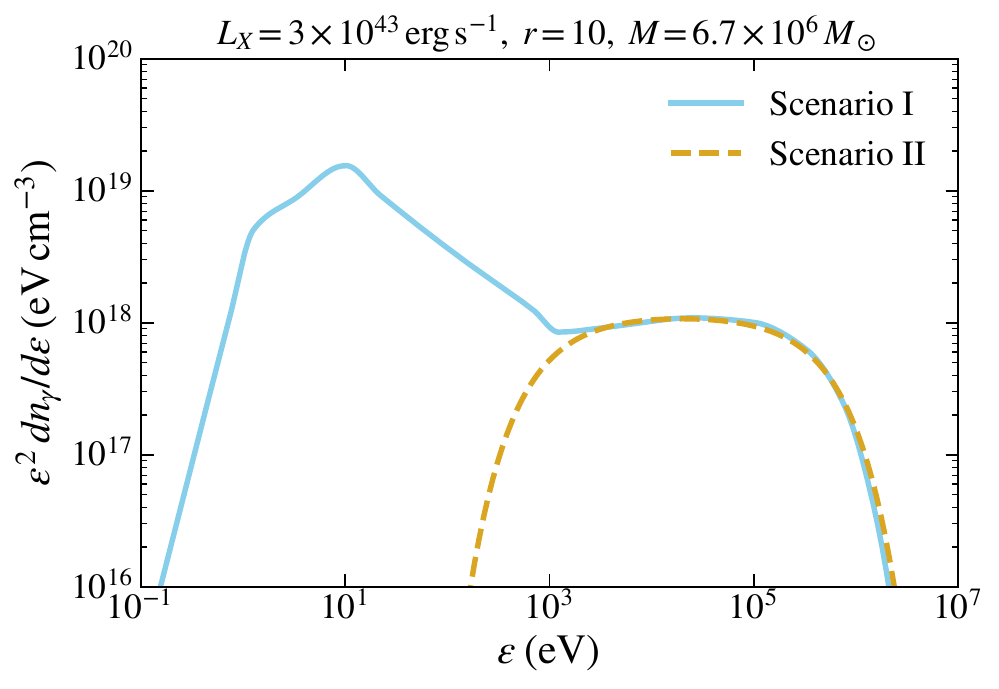}

\caption{
Differential number density of target photons
for scenario I, including both the coronal X-ray
and accretion-disk OUV components, and scenario II,
including only the coronal X-ray component.
Both distributions are shown for
$M=6.7 \cdot 10^6\,M_\odot$, $r=10$, and
$L_X=3 \cdot 10^{43}\,{\rm erg\,s^{-1}}$.
}
\label{fig:photon_field}
\end{figure}

The radiation in the corona is crucial to assess how quickly cosmic rays are cooled. We model the corona as a spherical region with a radius $\mathcal{R}_c$ as a fixed multiple of the gravitational radius of the black hole ($R_g$)

\begin{equation}
\mathcal{R}_c=r\mathcal{R}_g=\frac{GM}{c^2}r\simeq1.5\times10^{13}\left(\frac{r}{10}\right)\left(\frac{M}{10^7\,M_\odot}\right)\,{\rm cm}.
\label{eq:coronal_radius}
\end{equation}
We focus on NGC~1068 as a benchmark with a black hole mass $M=6.7\times 10^6\, M_\odot$~\cite{Padovani:2024ibi}. For the background radiation, we consider different scenarios:

\begin{itemize}
    \item \emph{scenario I}, where coronal X-rays coexist with the OUV photons from the accretion disk, leading to the characteristic big blue bump in the AGN spectra. We use the luminosity per unit energy $dL/d\varepsilon$ parameterized in Ref.~\cite{Marconi:2003tg}, including both the X-ray and the OUV component, normalizing it to a luminosity integrated between 2~keV and 10~keV, $L_X$, of $3\times 10^{43}\,\mathrm{erg/s}$~\cite{Padovani:2024ibi} as in Refs.~\cite{Fiorillo:2025ehn,Testagrossa:2026jcs}.

\item \emph{scenario II}, where only corona X-rays are included,
as in Ref.~\cite{Fiorillo:2023dts}.
The X-ray spectral luminosity is modeled as
\begin{equation}
\frac{dL}{d\varepsilon} = \mathcal{N}_X \frac{1}{\varepsilon} \exp\left(-\frac{\varepsilon}{\varepsilon_{\rm cut,high}}\right) \exp\left(-\frac{\varepsilon_{\rm cut,low}}{\varepsilon}\right),
\end{equation}
with $\varepsilon_{\rm cut,low}=0.8\,{\rm keV}$ and
$\varepsilon_{\rm cut,high}=500\,{\rm keV}$.
The normalization $\mathcal{N}_X$ is fixed by requiring $L_X=3\times10^{43}\,{\rm erg\,s^{-1}}$.
Indeed, if the OUV flux is produced significantly outside
of the corona, at radii of the order of
$r_{\rm OUV}\sim 75$ (see, e.g., Eq.~(A9) in
Ref.~\cite{Fiorillo:2023dts}, obtained from the
Shakura--Sunyaev theory of geometrically thin
accretion disks~\cite{1973A&A....24..337S,2012arXiv1203.6851M}),
the OUV field inside the corona is geometrically
suppressed by $(\mathcal{R}_c/r_{\rm OUV})^2$.

\end{itemize}
Thus, the two configurations bracket the uncertainties on the overall geometry of the system. The differential number density of target photons reads
\begin{equation}
n_{\rm back}(\varepsilon,r, L_X) = \frac{3} {4\pi \mathcal{R}_c^2\,c\,\varepsilon} \frac{dL}{d\varepsilon}(\varepsilon,L_X). \label{eq:background_photon_density}
\end{equation}
The resulting target photon densities for scenarios I and II are shown in Fig.~\ref{fig:photon_field} for the same coronal radius, $r=10$. This comparison isolates the effect of the additional OUV component on the photon field encountered by the accelerated particles.

\subsection{Particle acceleration and escape}
\label{subsec:timescales}

We consider two acceleration prescriptions: stochastic acceleration in a magnetized turbulent corona and particle acceleration in a magnetic-reconnection layer. These prescriptions are independent of the two radiation-field configurations defined above. The turbulent-corona model is studied with both photon-field scenarios I and II, whereas magnetic reconnection is considered with the X-ray-only photon field (scenario II). The two acceleration prescriptions also differ in the size and magnetization of the interaction region.

\paragraph*{Magnetized turbulent corona.}

In the turbulent scenario, CR protons are stochastically energized by magnetic fluctuations close to the central supermassive black hole (SMBH). The fundamental parameters describing the coronal plasma~\cite{Fiorillo:2024akm} are the Thomson optical depth $\tau_T$ and the plasma magnetization $\sigma$. Assuming charge neutrality, the thermal proton and electron number densities satisfy
\begin{equation}
n_p^{\rm gas}\simeq n_e=\frac{\tau_T}{\sigma_T\mathcal{R}_c}=\frac{\tau_T}{\sigma_T r\mathcal{R}_g}.
\label{eq:corona_density}
\end{equation}
The plasma magnetization and the corresponding magnetic-field strength are
\begin{equation}
\sigma=\frac{B^2}{4\pi n_p^{\rm gas}m_pc^2}.
\label{eq:corona_magnetization}
\end{equation}
\begin{equation}
B=\left(4\pi\sigma n_p^{\rm gas}m_pc^2\right)^{1/2}.
\label{eq:magnetic_field_turbulent}
\end{equation}

To characterize the turbulent component of the magnetic field, we introduce $\sigma_{\rm tur}=\eta_B\sigma$, where $\eta_B=(\delta B/B)^2$ quantifies the relative strength of magnetic fluctuations. The turbulence coherence length is $\ell=\eta_{\rm coh}\mathcal{R}_c$. In this framework, proton acceleration is primarily driven by non-resonant interactions with large-scale turbulent magnetic structures. The characteristic acceleration time is energy-independent and reads~\cite{Fiorillo:2024akm,Comisso:2019frj}
\begin{equation}
\tau_{\rm acc}=\frac{10\ell}{\sigma_{\rm tur}c}=\frac{10\eta_{\rm coh}\mathcal{R}_c}{\sigma_{\rm tur}c}.
\label{eq:tacc_turbulent}
\end{equation}
At the same time, turbulence controls particle confinement, with a time given by~\cite{Kempski:2023ikw,Comisso:2024ymy,Fiorillo:2024akm}
\begin{equation}
\tau_{\rm esc,p}(E)=\frac{\mathcal{R}_c}{c}\max\!\left[1,\frac{1}{\eta_{\rm coh}}\left(\frac{eB\ell}{E}\right)^{1/3}\right].
\label{eq:tesc_turbulent}
\end{equation}
This prescription ensures that the diffusive confinement always satisfies $\tau_{\rm esc,p}\geq \mathcal{R}_c/c$.

For our reference configuration, we adopt $\tau_T=0.5$, $\sigma=1$,  $\eta_B=1$, $\eta_{\rm coh}=0.5$, and $r=10$, motivated by the parameters inferred from NGC~1068 neutrino emission~\cite{Fiorillo:2024akm,Testagrossa:2026jcs}. These assumptions yield $n_p^{\rm gas}\simeq7.6\times10^{10}\,{\rm cm^{-3}}$, $B\simeq3.8\times10^4\,{\rm G}$, and $\tau_{\rm acc}\simeq1.65\times10^3\,{\rm s}$. 

\paragraph*{Magnetic reconnection.}

In this scenario, non-thermal protons are accelerated through the rapid conversion of magnetic energy into particle kinetic energy within a compact reconnection layer. Following Ref.~\cite{Fiorillo:2023dts}, we determine the magnetic field by assuming that a fraction $\eta_X$ of the incoming Poynting flux is converted into the total coronal X-ray luminosity,
\begin{equation}
\frac{L_{X,\rm tot}}{4\pi\mathcal{R}_c^2}=\eta_X\beta_{\rm rec}\frac{cB^2}{4\pi}.
\label{eq:reconnection_Xray_luminosity}
\end{equation}
Here $\beta_{\rm rec}c$ is the characteristic inflow speed into the reconnection region, giving
\begin{equation}
B=\left(\frac{L_{X,\rm tot}}{\eta_X\beta_{\rm rec}c\mathcal{R}_c^2}\right)^{1/2}.
\label{eq:B_reconnection}
\end{equation}
The bolometric X-ray luminosity $L_{X,\rm tot}\simeq4L_X$ is obtained by integrating the intrinsic X-ray spectrum over the adopted photon-energy range. In our reference configuration we use $r=2$, $\beta_{\rm rec}=0.1$, and $\eta_X=0.5$, leading to $B\simeq1.4\times10^5\,{\rm G}$~\cite{Fiorillo:2023dts}.

The characteristic acceleration rate in the upstream reconnection region is~\cite{Fiorillo:2023dts}
\begin{equation}
\frac{dE}{dt}\simeq\beta_{\rm rec}eBc.
\label{eq:reconnection_acceleration_rate}
\end{equation}
So, the corresponding acceleration time reads
\begin{equation}
\tau_{\rm acc}(E)\simeq\frac{E}{\beta_{\rm rec}eBc},
\label{eq:tacc_reconnection}
\end{equation}
which is much faster than the other relevant dynamical processes. Differently from the magnetized-turbulent scenario, here, we assume a pair-dominated plasma, with a proton density much lower than the electron density. Therefore, 
once injected into the radiative region, protons are no longer actively accelerated and escape ballistically,
\begin{equation}
\tau_{\rm esc,p}=\frac{\mathcal{R}_c}{c}.
\label{eq:tesc_reconnection}
\end{equation}

\subsection{Energy-loss mechanisms}
\label{subsec:hadronic_losses}

High-energy CRs propagating through the coronal environment undergo several energy-loss processes, including synchrotron radiation, BH pair production, and photopion interactions with the ambient radiation field. In the magnetized turbulent-corona scenario, CRs additionally undergo hadronuclear interactions with the thermal proton population. The treatment of these processes is particularly important for determining the steady-state particle distributions. In this work, we compare two different prescriptions: the conventional CEL approximation and the TD approach. While synchrotron and Bethe--Heitler losses are always treated as continuous processes, photohadronic and hadronuclear interactions are described either through their average energy losses or as discrete collisions involving nucleon regeneration and charge exchange.

\subsubsection{Continuous-energy-loss approximation}
\label{sec:continuous_transport}

In the conventional CEL approximation, energy losses are described by a continuous energy drift, determined by the average energy transferred from the parent particle to secondary particles. The corresponding operator to the steady-state proton transport equation is
\begin{equation}
\mathcal{L}_p^{\rm CEL}(E)=-\frac{d}{dE}\left[b_p^{\rm CEL}(E)n_p(E)\right],
\label{eq:cel_collision_operator}
\end{equation}
where $n_p(E)$ is the differential proton number density and $b_p^{\rm CEL}(E)>0$ denotes the total energy-loss rate, defined as
\begin{equation}
b_p^{\rm CEL}(E)\equiv-\frac{dE}{dt}=E\left[\tau_{\rm syn,p}^{-1}(E)+\tau_{\rm BH}^{-1}(E)+\tau_{p\gamma,\rm loss}^{-1}(E)+\tau_{pp,\rm loss}^{-1}(E)\right] = E \cdot \tau^{-1}_{\rm cool, p},
\label{eq:continuous_loss_rate}
\end{equation}
where $\tau_{\rm syn,p}(E)$, $\tau_{\rm BH}(E)$, $\tau_{p\gamma,\rm loss}(E)$, $\tau_{pp,\rm loss}(E)$ are the synchrotron, BH pair production, photopion and hadronuclear collisions cooling timescales, respectively; $\tau_{\rm cool, p}$, instead, represents the total cooling proton timescale.
As said above, the hadronuclear collision contribution is  included only for the magnetized turbulent-corona scenario.
The proton synchrotron cooling timescale is
\begin{equation}
\tau_{\rm syn,p}(E)=\frac{3m_p^4c^3}{4\sigma_Tm_e^2u_BE},
\label{eq:tsyn_proton}
\end{equation}
where $u_B=\frac{B^2}{8\pi}$ is the magnetic energy density. Although this expression is employed in both acceleration scenarios, the different magnetic-field strengths lead to different synchrotron cooling timescales. The BH pair production timescale follows from the prescription of Ref.~\cite{1992ApJ...400..181C}, using the background photon distribution adopted for each configuration.

For photopion interactions, we use the Chromo open code with the SOPHIA 2.0 event generator~\cite{Fedynitch:2025mgj,Mucke:1999yb} to extract cross sections, branching probabilities, and microscopic leading-nucleon energy fractions useful to calculate the overall timescale. 
In particular, we start by defining the total photopion interaction rate, namely the frequency with which a given nucleon $a={p,n}$ interacts with an isotropic photon background field~\cite{Aloisio:2017iyh}

\begin{equation}
\Gamma_{a\gamma}(E)\equiv\tau_{a\gamma,\rm int}^{-1}(E)=\frac{c}{2\gamma_a^2}\int_{\varepsilon'_{\rm th}}^{\infty}d\varepsilon'\,\sigma_{a\gamma}(\varepsilon')\varepsilon'J\!\left(\frac{\varepsilon'}{2\gamma_a}\right),
\label{eq:photopion_interaction_rate}
\end{equation}
where $\gamma_a=E/(m_ac^2)$ is the Lorentz factor of the incident nucleon, $\varepsilon'$ denotes the photon energy in its rest frame, $\varepsilon'_{\rm th}$ is the pion-production threshold, and $J$ accounts for the background photon distribution,
\begin{equation}
J(y)=\int_y^\infty d\varepsilon\,\frac{n_{\rm back}(\varepsilon)}{\varepsilon^2}.
\label{eq:td_photon_integral}
\end{equation}
The corresponding photopion energy-loss timescale is
\begin{equation}
\tau_{a\gamma,\rm loss}^{-1}(E)=\Gamma_{a\gamma}(E)K_{a\gamma,\rm loss}(E),
\label{eq:tloss_photopion}
\end{equation}
where $K_{a\gamma,\rm loss}(E)$ is the effective inelasticity, averaged over the background photon field, whose typical values are of the order of $20-50\%$. Details about the extraction of microscopic processes and the construction of the corresponding effective quantities are given in App.~\ref{sec:app_photopion}. Here, to get a qualitative handle on the interaction properties, we note that the cross section $\sigma_{N\gamma}(\varepsilon')$ peaks close to its threshold, so a nucleon with energy $E$ interacts primarily with target photons with energy $\varepsilon_\gamma \sim 0.3\,\mathrm{GeV}^2/E$. Therefore, if the target photon spectrum $n_{\rm back}$ decreases faster than $\varepsilon^{-1}$, the amount of photohadronic target increases with nucleon energy $E$, leading to a larger $\Gamma_{N\gamma}$ (see e.g. Ref.~\cite{Fiorillo:2021hty}).
For hadronuclear collisions, we analogously use Chromo to calculate cross sections, branching fractions, and inelasticities employing the Sibyll 2.3e interaction model~\cite{Fedynitch:2025mgj,Ahn:2009wx,Fedynitch:2018cbl,Riehn:2019jet}. In particular, the associated energy-loss timescale satisfies 
\begin{equation}\label{eq:hadronuclear_loss_times}
    \tau_{ap,\rm loss}^{-1}(E)=K_{ap}^{\rm eff}\Gamma_{ap}(E), 
\end{equation}
where $K_{ap}^{\rm eff}$ denotes the effective hadronuclear inelasticity for  the nucleon $a$. 
By contrast, $\Gamma_{ap}(E) = n_p^{\rm gas}c\sigma_{ap}^{\rm inel}(E)$ is the interaction rate for hadronuclear collisions between the nucleon $a$ and the proton gas. Further details are provided in Appendix~\ref{sec:app_hadronuclear}. Here, we note that the hadronuclear cross sections are roughly energy-independent for energies larger than a few GeV, with a typical order of magnitude $\sigma^{\rm inel}\sim 40\, \rm mb$. 

In the CEL approximation, these interaction rates contribute only through their average energy losses. Consequently, individual collisions are not followed explicitly, and neither the discrete redistribution of nucleon energies nor proton--neutron conversion is taken into account. In particular, only the proton distribution is evolved in this treatment.

\subsubsection{Discrete interactions: the trickle-down approach}
\label{sec:trickledown}

The TD approach~\cite{Unger:2015laa} goes beyond the CEL approximation by treating photohadronic and hadronuclear interactions as discrete processes. Each inelastic collision removes a parent nucleon from its original energy bin and reinjects a leading proton or neutron at a lower energy, depending on the interaction channel and its inelasticity. This treatment explicitly accounts for nucleon regeneration, energy redistribution, and charge-exchange processes.
Synchrotron and BH losses are always treated as continuous mechanisms, with the energy-loss rate modified as
\begin{equation}
b_p^{\rm TD}(E)=E\left[\tau_{\rm syn,p}^{-1}(E)+\tau_{\rm BH}^{-1}(E)\right].
\label{eq:td_continuous_loss}
\end{equation}
On the other hand, inelastic photohadronic and hadronuclear collisions are described through explicit nucleon-removal and secondary-injection terms. For a nucleon $a\in\{p,n\}$, the corresponding collision operator is
\begin{equation}
\mathcal{C}_a^{\rm TD}(E)=\Gamma_a^{\rm int}(E)n_a(E)-\sum_{b\in\{p,n\}}Q_{a\leftarrow b}^{\rm TD}(E),
\label{eq:td_collision_operator}
\end{equation}
where $n_a (E)$ denotes either the non-thermal proton or the neutron differential number densities, and the total interaction rates sum both the hadronic and the photohadronic collisions: 
\begin{equation}
\Gamma_p^{\rm int}(E)=\Gamma_{p\gamma}(E)+\Gamma_{pp}(E),\qquad\Gamma_n^{\rm int}(E)=\Gamma_{n\gamma}(E)+\Gamma_{np}(E).
\label{eq:td_proton_interaction_rate}
\end{equation}
The term $Q_{a\leftarrow b}^{\rm TD}(E)$ describes the production of a daughter nucleon $a$ from a parent nucleon $b$,
\begin{equation}
Q_{a\leftarrow b}^{\rm TD}(E)=\sum_{j\in\{b\gamma,bp\}}Q_{a\leftarrow b}^{(j)}(E).
\label{eq:td_total_reinjection}
\end{equation}

In order to construct these secondary injection terms, we use the  mean-inelasticity approximation, where the daughter energy $E$ is related to the parent energy $E_*$ by
\begin{equation}
E=\xi_{a\rightarrow b}^{(j)}(E_*)E_*,\qquad\xi_{a\rightarrow b}^{(j)}(E_*)=1-k_{a\rightarrow b}^{(j)}(E_*),
\label{eq:td_energy_mapping}
\end{equation}
where $k_{a\rightarrow b}^{(j)}$ is the effective channel inelasticity, while $\xi_{a\rightarrow b}^{(j)}(E)$ is the mean fraction of the parent energy retained by the leading daughter nucleon.
The secondary injection term is then
\begin{equation}
Q_{b\leftarrow a}^{(j)}(E)=\frac{\Gamma_{a\rightarrow b}^{(j)}(E_*)n_a(E_*)}{\left|\xi_{a\rightarrow b}^{(j)}(E_*)+E_*\,d\xi_{a\rightarrow b}^{(j)}(E_*)/dE_*\right|}.
\label{eq:td_injection}
\end{equation}
Here $E_*$ is obtained by inverting the energy mapping. The denominator is the Jacobian of the parent-to-daughter energy transformation, ensuring the correct redistribution of particles in differential energy space. $\Gamma_{a\rightarrow b}^{(j)}(E)$ represents the partial interaction rate, namely  the total interaction rate weighted by the probability of producing  the leading nucleon $b$. The specific implementations of $\Gamma_{a\rightarrow b}^{(j)}$ and $\xi_{a\rightarrow b}^{(j)}$ for photohadronic and hadronuclear interactions are reported in Apps.~\ref{sec:app_photopion} and~\ref{sec:app_hadronuclear}, respectively.

In addition to inelastic interactions, neutrons may escape from the corona or decay into protons. Since neutrons are not magnetically confined, their escape timescale is taken to be the ballistic crossing time, $\tau_{\rm esc,n}=\mathcal{R}_c/c$. Their Lorentz-dilated beta-decay timescale is
\begin{equation}
\tau_{\rm dec,n}(E)=\tau_{n,0}\frac{E}{m_nc^2},\qquad\tau_{n,0}\simeq880\,{\rm s}.
\label{eq:tdec_neutron}
\end{equation}
The corresponding proton injection from neutron decay is approximated as
\begin{equation}
Q_{p\leftarrow n}^{\rm dec}(E)\simeq\frac{n_n(E)}{\tau_{\rm dec,n}(E)},
\label{eq:neutron_decay_injection}
\end{equation}
neglecting the small energy difference between the parent neutron and the daughter proton.

\section{Cosmic-ray evolution equations}
\label{sec:cr_evolution}

Having specified the acceleration, escape, and collision prescriptions, in this section, we write the complete steady-state CR transport equations.  We first present the canonical CEL treatment and then discuss the TD equations. Numerical discretization and boundary conditions used to solve the equations are reported  in App.~\ref{app:numerical_transport}.

\subsection{Continuous-energy-loss transport}

The canonical steady-state proton distribution in corona environments can be obtained from a Fokker--Planck~(FP) equation~\cite{Fiorillo:2024akm}

\begin{equation}
-\frac{d}{dE}\left[\frac{E^4}{\tau_{\rm acc}(E)}\frac{d}{dE}\left(\frac{n_p(E)}{E^2}\right)\right]-\frac{d}{dE}\left[b_{p}^{\rm CEL}(E)n_p(E)\right]+\frac{n_p(E)}{\tau_{\rm esc,p}(E)}=Q_{\rm inj}(E),
\label{eq:fp_continuous}
\end{equation}
where each term represents, from left to right, the acceleration operator, the collisional energy loss term, the escape and injection rates.

For the magnetized-turbulent scenario,  stochastic particle acceleration redistributes the injected particles over a broad energy range, so the high-energy steady-state spectrum is insensitive to the detailed shape of the low-energy seed population. We inject a narrow proton spectrum as,
\begin{equation}
Q^{\rm MT}_{\rm inj}(E)=Q_{\rm seed,0}E^2\exp\!\left[-\frac{1}{2}\left(\frac{\log_{10}(E/E_0)}{\sigma_{\rm dex}}\right)^2\right],
\label{eq:seed_injection}
\end{equation}
with $\sigma_{\rm dex}=0.12$ and $E_0=20\,{\rm GeV}$. The $E^2$ prefactor can be interpreted as the phase-space weighting of an isotropic relativistic particle distribution in the momentum space, while the Gaussian factor in logarithmic energy confines the injected particles to a narrow energy range. The injection spectrum peaks slightly above $E_0$, at approximately
$23\,{\rm GeV}$.
We have also verified that varying the low-energy numerical setup and the seed-injection parameters, results are unchanged.
Modeling the acceleration in this way is only an approximated representation of the energization process, as discussed in Refs.~\cite{Bresci:2022awc,Lemoine:2022rpj}, but is expected to describe well the qualitative proton spectrum~\cite{Fiorillo:2024akm}.

For the magnetic reconnection case, instead, the acceleration rate is so high that  we  do not model it as an explicit term in the transport equation for the radiative region. Following Ref.~\cite{Fiorillo:2023dts}, we inject a broken-power-law proton spectrum

\begin{equation}
Q^{\rm MR}_{\rm inj}(E)=Q_{\rm MR,0}\begin{cases}(E/E_{\rm br})^{-1}, & E\leq E_{\rm br},\\ (E/E_{\rm br})^{-s}, & E>E_{\rm br},\end{cases}
\label{eq:reconnection_injection}
\end{equation}
where we adopt $s=3$, assuming a strong guide field~\cite{Fiorillo:2023dts}. The characteristic break energy is related to the proton magnetization, $E_{\rm br}\sim\sigma_p m_pc^2$, where $\sigma_p=B^2/(4\pi n_pm_pc^2)$. We adopt $\sigma_p=2.66\times10^4$, corresponding to $E_{\rm br}\simeq25\,\mathrm{TeV}$, following Ref.~\cite{Fiorillo:2023dts}. 

\subsection{Coupled proton--neutron trickle-down transport}

In the TD approach, the steady-state proton and neutron equations are coupled. For the protons, the equation becomes

\begin{equation}
\begin{aligned}
&-\frac{d}{dE}\left[\frac{E^4}{\tau_{\rm acc}(E)}\frac{d}{dE}\left(\frac{n_p(E)}{E^2}\right)\right]
-\frac{d}{dE}\left[b_p^{\rm TD}(E)n_p(E)\right] +\left[\tau_{\rm esc,p}^{-1}(E)+\Gamma_p^{\rm int}(E)\right]n_p(E) = \\
&=Q_{\rm inj}(E)+Q_{p\leftarrow p}^{\rm TD}(E)+Q_{p\leftarrow n}^{\rm TD}(E)+Q_{p\leftarrow n}^{\rm dec}(E).
\end{aligned}
\label{eq:td_proton_transport}
\end{equation}
The steady-state neutron equation is given, instead, by
\begin{equation}
[\tau_{\rm esc,n}^{-1}(E)+\tau_{\rm dec,n}^{-1}(E)+\Gamma_n^{\rm int}(E)]n_n(E)= Q_{n\leftarrow p}^{\rm TD}+Q_{n\leftarrow n}^{\rm TD}.
\label{eq:td_neutron_transport}
\end{equation}

\paragraph*{Normalization of the particle distributions.}

We employ two different physically motivated normalizations for the acceleration prescriptions. For the magnetized turbulent corona, we follow the energy-budget arguments of Ref.~\cite{Testagrossa:2026jcs}, where the ratio between the non-thermal proton energy density~$(u_p)$ and the turbulent magnetic energy density~$(u_{B,\rm tur})$ is related to the fraction of magnetic dissipation power transferred to non-thermal protons as

\[
\frac{u_p}{u_{B,\rm tur}}
=
\frac{5}{2}
\frac{\eta_{\rm rec}}{\sqrt{\sigma_{\rm tur}}}
\frac{L_p}{L_B},
\]

where \(L_p\) is the power transferred to non-thermal protons, \(L_B\) is the turbulent magnetic dissipation power, and \(\eta_{\rm rec}\) is the magnetic reconnection rate. Adopting \(\eta_{\rm rec}=0.1\)~\cite{Testagrossa:2026jcs}, and the maximal proton acceleration efficiency \(L_p/L_B=1\), we obtain

\[
u_p\equiv\int_{E_{\min}}^{E_{\max}}dE\,E\,n_p(E)
=0.25\,u_{B,\rm tur}.
\]

This normalization follows from the assumption
of fast magnetic reconnection in a strongly
magnetized turbulent plasma, adopting the
approximate Alfvén-speed prescription of
Ref.~\cite{Testagrossa:2026jcs}. We consider this condition as a benchmark and  impose the same energy-density normalization for both the CEL and TD solutions to isolate the impact of the different energy-loss treatments.

For magnetic reconnection, we normalize the injection spectrum by assuming that a fraction $\eta_p^\ast=0.1$ of the magnetic energy budget is supplied to non-thermal protons over the coronal crossing time~\cite{Fiorillo:2023dts},
\begin{equation}
\int_{E_{\min}}^{E_{\max}}dE\,E\,Q^{\rm MR}_{\rm inj}(E)=\eta_p^\ast\frac{cu_B}{\mathcal{R}_c}.
\label{eq:normalization_reconnection}
\end{equation}
The same injected spectrum and normalization are adopted for the CEL and TD calculations, with no subsequent rescaling of the steady-state distributions. Therefore, the comparison between continuous and TD transport is performed at fixed steady-state proton energy density in the turbulent corona, but at fixed injected proton power in magnetic reconnection. These normalizations allow us to study the effect of the particle transport in the two approaches.

\section{Neutrino production rates}
\label{subsec:neutrino_production}

Equipped with the steady-state proton and neutron distributions, we compute the differential
neutrino production rates  from photohadronic and hadronuclear interactions. While photohadronic neutrino production is considered in both acceleration scenarios, hadronuclear contributions are included only in the magnetized turbulent-corona model.
The neutrino production rate from photohadronic interactions is estimated following the analytical parametrizations of Ref.~\cite{Kelner:2008ke}.
For an isotropic target photon field, the differential production rate of neutrinos of
flavor $\ell$ is
\begin{equation}
Q_{\nu_\ell}^{p\gamma}(E_\nu)
=
c\int_{E_\nu}^{\infty}
\frac{dE_p}{E_p}\,
n_p(E_p)
\int_{\varepsilon_{\min}(E_p)}^{\infty}
d\varepsilon\,
n_{\rm back}(\varepsilon)\,
\Phi_{\nu_\ell}(\eta,x),
\label{eq:neutrino_photohadronic}
\end{equation}
where $n_p(E_p)$ is the differential steady-state
proton distribution and $n_{\rm back}(\varepsilon)$
is the differential number density of target
photons. The dimensionless kinematic variables are
\begin{equation}
\eta =
\frac{4\varepsilon E_p}{m_p^2c^4},
\qquad
x = \frac{E_\nu}{E_p},
\label{eq:neutrino_kinematics}
\end{equation}
and the lower integration boundary is determined
by the threshold for pion production,
\begin{equation}
\varepsilon_{\min}(E_p)
=
\frac{\eta_0 m_p^2c^4}{4E_p},
\qquad
\eta_0 =
2\frac{m_\pi}{m_p}
+
\frac{m_\pi^2}{m_p^2}.
\label{eq:neutrino_photon_threshold}
\end{equation}
The functions $\Phi_{\nu_\ell}(\eta,x)$ describe
the energy distributions of secondary neutrinos,
including contributions from the decay of
secondary mesons and muons.

In the TD treatment, both proton and neutron
populations contribute to neutrino production.
For neutron-induced photohadronic interactions,
we adopt the same neutrino production kernel
as for protons, assuming approximate isospin
symmetry. Accordingly, the total photohadronic
neutrino production rate is approximated as
\begin{equation}
Q_{\nu_\ell}^{\gamma N}(E_\nu)
\simeq
Q_{\nu_\ell}^{p\gamma}(E_\nu)
+
Q_{\nu_\ell}^{n\gamma}(E_\nu),
\label{eq:neutrino_photohadronic_total}
\end{equation}
where the neutron contribution is evaluated
by replacing the proton distribution with
the corresponding neutron distribution in
Eq.~\eqref{eq:neutrino_photohadronic}.

For the magnetized turbulent-corona scenario,
we additionally consider neutrino production from inelastic collisions of relativistic
nucleons with the thermal proton population. The corresponding differential production rate
is given by~\cite{Koldobskiy:2021nld}
\begin{equation}
Q_{\nu_\ell}^{Np}(E_\nu)
=
c\,n^{\rm gas}_{\rm p}
\sum_{N=p,n}
\int_{E_\nu}^{\infty}
dE_N\,n_N(E_N) \frac{d\sigma_{Np\rightarrow\nu_\ell}}
{dE_\nu}(E_N,E_\nu),
\label{eq:neutrino_hadronuclear}
\end{equation}
where  $d\sigma_{Np\rightarrow\nu_\ell}/dE_\nu$
is the inclusive differential cross section
for neutrino production.

We evaluate the hadronuclear neutrino spectra using the AAfrag parametrizations discussed in
Ref.~\cite{Koldobskiy:2021nld}, which provides the energy
distributions of secondary neutrinos produced in inelastic proton--proton collisions.

For neutron--proton interactions, we approximate the inclusive neutrino production cross section by its proton--proton counterpart,
\begin{equation}
\frac{d\sigma_{np\rightarrow\nu_\ell}}{dE_\nu}
\simeq
\frac{d\sigma_{pp\rightarrow\nu_\ell}}{dE_\nu}.
\label{eq:neutrino_np_approximation}
\end{equation}

This approximation is motivated by the approximate isospin symmetry of hadronic interactions and the similarity of inclusive pion production
at high energies. Furthermore, since neutron-induced interactions provide a subdominant contribution to the total all-flavour neutrino emission compared to proton-induced interactions, we consider these approximations acceptable for the goal of our analysis.

This allows us to consistently include neutrino production from the neutron population generated in the TD treatment. Therefore, the total neutrino production rate is
\begin{equation}
Q_{\nu_\ell}^{\rm tot}(E_\nu)
=
Q_{\nu_\ell}^{\gamma N}(E_\nu)
+
Q_{\nu_\ell}^{Np}(E_\nu),
\label{eq:neutrino_total_production}
\end{equation}
Throughout this work, we consider the all-flavor neutrino production rate, obtained by summing over neutrinos and antineutrinos of all flavors,
\begin{equation}
Q_{\nu}^{\rm all}(E_\nu)
=
\sum_{\ell=e,\mu,\tau}
\left[
Q_{\nu_\ell}^{\rm tot}(E_\nu)
+
Q_{\bar{\nu}_\ell}^{\rm tot}(E_\nu)
\right].
\label{eq:neutrino_all_flavor}
\end{equation}

This approach, however, assumes that secondary particles, such as pions and muons produced in photohadronic and hadronuclear interactions, decay without undergoing significant energy losses. Given the strong magnetic fields expected in AGN coronae, synchrotron cooling may affect this assumption. For charged pions, the synchrotron cooling timescale becomes comparable to the decay timescale at energies of approximately $2.9 \, \rm PeV$ and $0.76\, \rm PeV$ for the magnetized turbulent corona and magnetic reconnection scenarios, respectively. Therefore, pion cooling can be safely neglected over the energy range relevant to the main neutrino emission. For muons, instead, the corresponding critical energies are approximately $156\, \rm  TeV$ and $41\, \rm  TeV$. Since neutrinos from muon decay typically carry about one-third of the parent muon energy, synchrotron cooling may become relevant for neutrino energies approaching approximately $50 \, \rm TeV$ and $14 \, \rm TeV$ in the two cases, respectively. As the predicted neutrino spectra peak below these characteristic energies, the prompt-decay approximation is expected to provide a reasonable description of the main emission features. Nonetheless, in App.~\ref{app:secondary_cooling}, we assess the impact of secondary cooling including synchrotron losses of charged pions and muons. The results show that neutrino fluxes are modified only after the peak, without changing our conclusions about the differences between the TD and CEL treatments.

\section{Magnetized turbulent corona: numerical results}~\label{Sec:results_cosmic_rays_magnetized}

We now compare the CEL  and TD treatments for a magnetized turbulent corona, considering both prescriptions for the target photon field. We start by discussing the timescales, and then we show the difference in the CR number densities and in the neutrino production rates.

\subsection{Physical timescales and neutron escape}

\begin{figure*}[h!]
    \centering
    \includegraphics[width=0.48\textwidth]
    {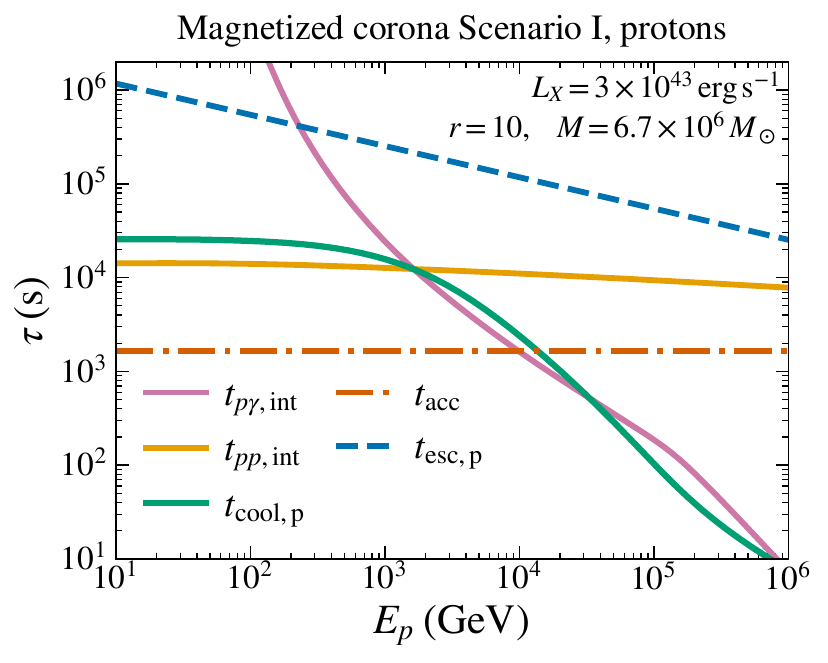}
    \includegraphics[width=0.48\textwidth]
    {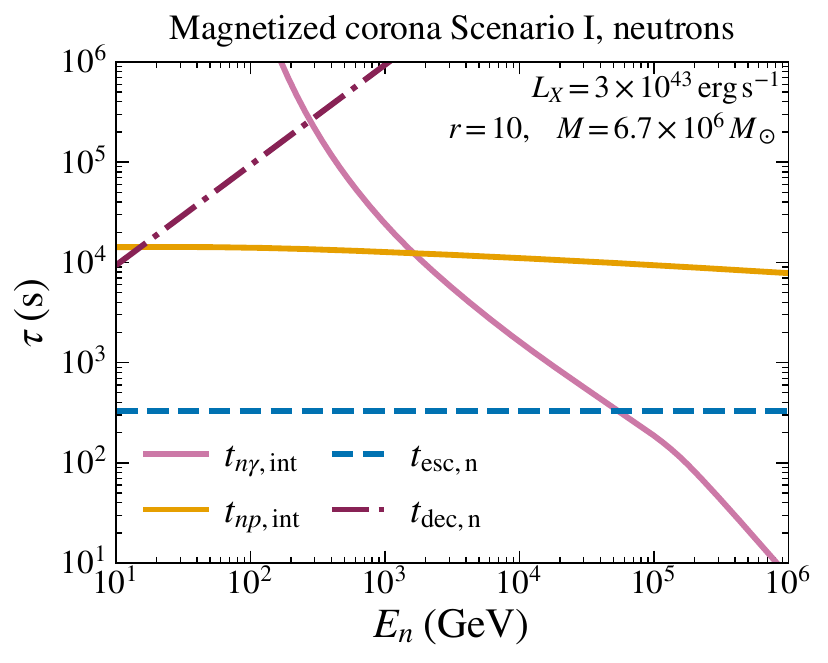}
    \caption{
    Characteristic timescales as a function of particle energy  for protons (left panel) and neutrons (right panel) in  Scenario I of the magnetized turbulent corona.
    The left panel shows the proton photohadronic and
    hadronuclear interaction timescales, together with
    the total cooling, acceleration, and escape timescales.
    The right panel shows the neutron photohadronic and
    hadronuclear interaction timescales, together with
    the escape and decay timescales.
    The adopted parameters are
    $L_X=3\times10^{43}\,\mathrm{erg\,s^{-1}}$,
    $r=10$, and $M=6.7\times10^6\,M_\odot$.
    }
    \label{fig:timescales_scenario_I}
\end{figure*}

\begin{figure*}[t]
    \centering
    \includegraphics[width=0.48\textwidth]
    {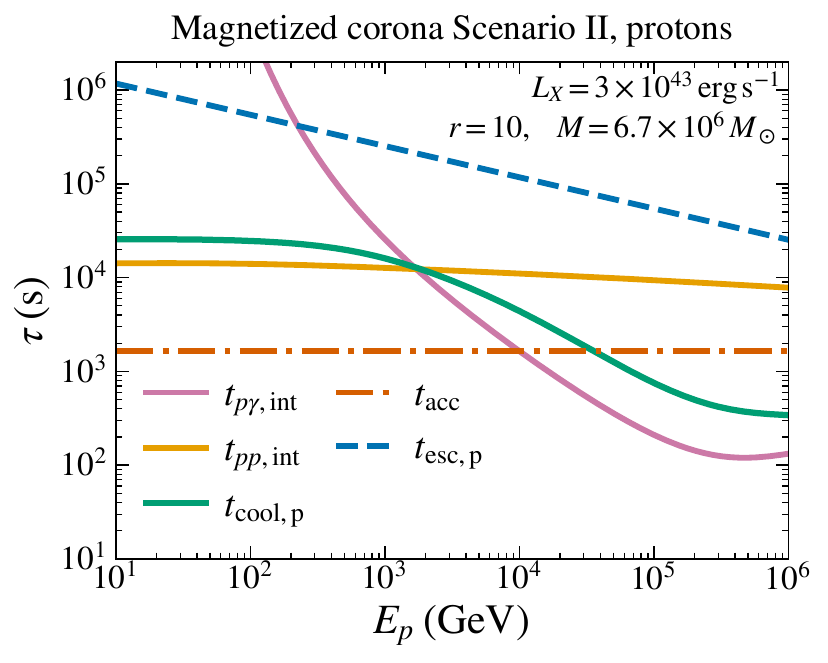}
    \includegraphics[width=0.48\textwidth]
    {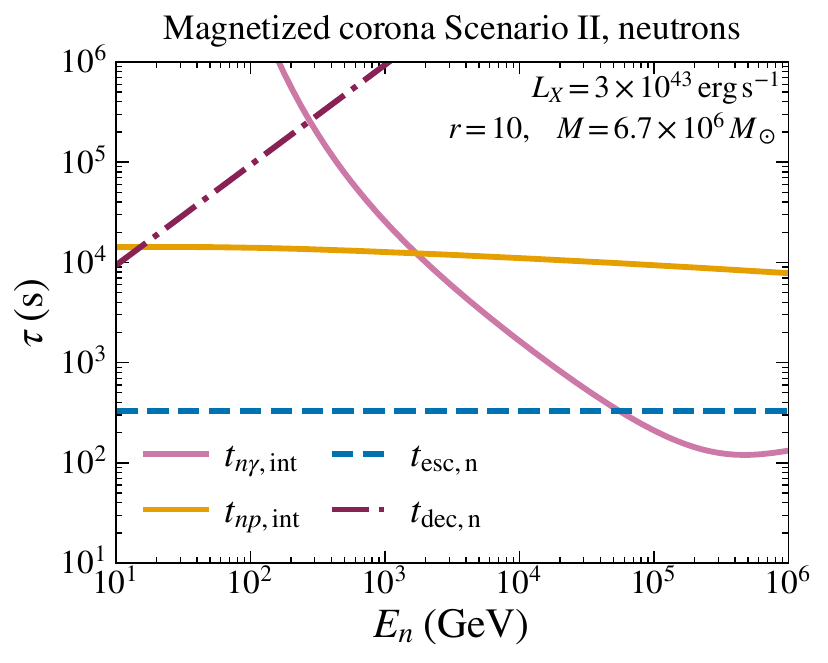}
    \caption{
    Same as Fig.~\ref{fig:timescales_scenario_I},
    but for scenario II, in which the background
    photon field contains only the coronal X-ray
    component.
    }
    \label{fig:timescales_scenario_II}
\end{figure*}

Figures~\ref{fig:timescales_scenario_I} and~\ref{fig:timescales_scenario_II} show the characteristic timescales for scenarios I and II, respectively. In both cases, stochastic acceleration takes place on an approximately energy-independent timescale, $\tau_{\mathrm{acc}}\simeq 1.65\times10^3\,\mathrm{s}$, whereas diffusive proton escape is considerably slower. Neutrons, instead, escape ballistically on the light-crossing time for our benchmark corona. In the relevant energy range, their escape is nearly instantaneous in comparison with their decay or photohadronic interactions. So proton-to-neutron conversion opens a much faster escape route for particles that would otherwise remain magnetically confined; this route is artificially excluded in the CEL treatment.

In addition to making protons escape faster, the TD approach also quantitatively alters the proton interaction rate, since the photohadronic and hadronic interactions leading to proton-to-neutron conversion are neglected in the CEL approach. The quantitative importance of this effect depends on what limits proton acceleration. In scenario I (Fig. 2), the additional optical–UV radiation from the accretion disk enhances BH cooling, which contributes substantially to the total proton energy losses. Consequently, the continuous cooling time already becomes comparable to the acceleration time around the energy at which the proton distribution turns over. In scenario II (Fig. 3), the optical–UV component is absent and BH losses are less important. Photopion interactions then play the key role in limiting proton acceleration: around $E_p\sim10^4\,\mathrm{GeV}$, their interaction time approaches the acceleration time while remaining shorter than the corresponding continuous energy-loss time. Thus, in scenario II, we anticipate a larger correction from the TD approach to the maximum energy reached by non-thermal protons.

This distinction matters because an inelastic collision does more than remove the average energy fraction accounted for by continuous cooling. It removes the parent nucleon from its original energy bin and reinjects a leading nucleon at lower energy; when the daughter is a neutron, it can subsequently escape. Hence, we expect vast changes in the proton residence time in both scenarios, but a more pronounced shift of the high-energy cutoff in scenario II, where photohadronic interactions are the principal process competing with acceleration.

\subsection{Steady-state nucleon distributions and injection requirements}

\begin{figure*}[t]
    \centering
    \includegraphics[width=0.48\textwidth]
    {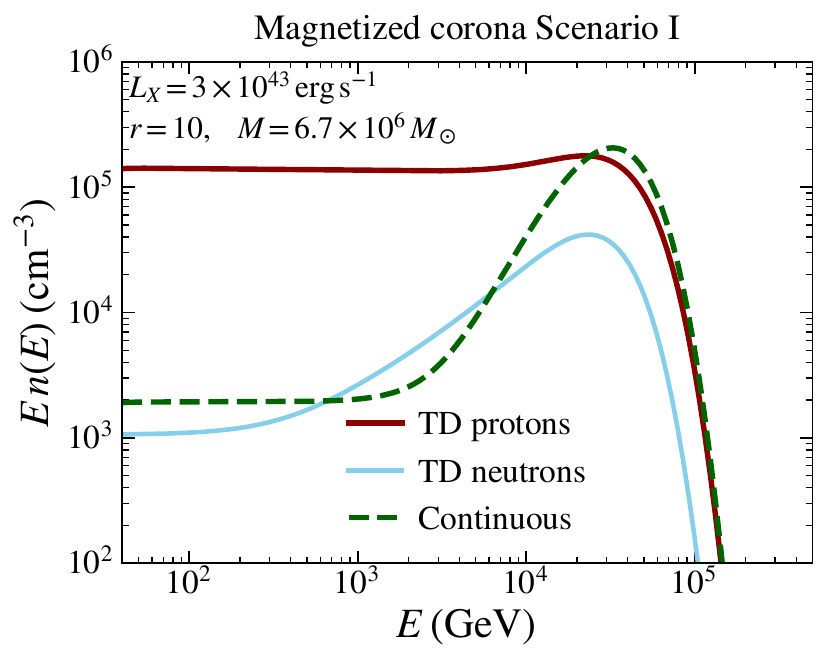}
    \includegraphics[width=0.48\textwidth]
    {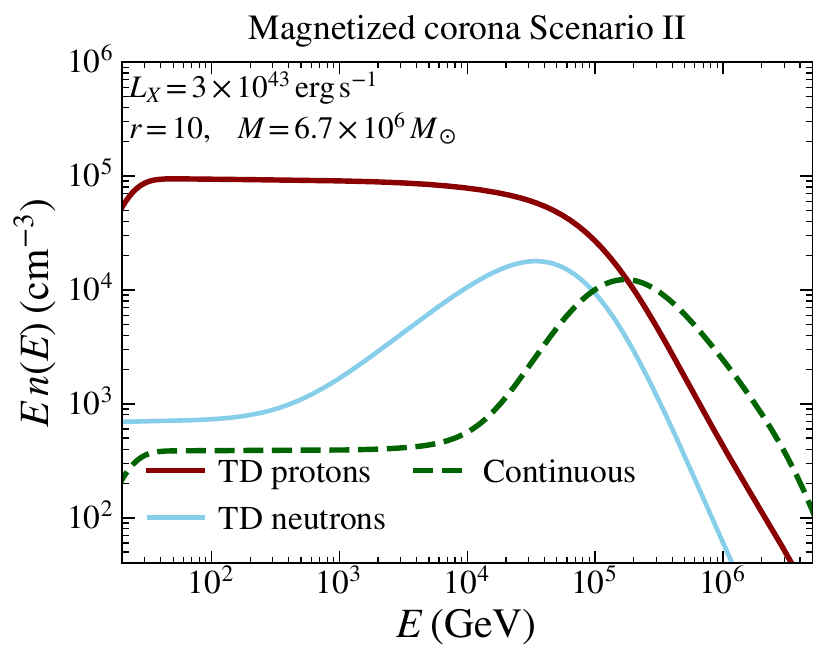}
    \caption{
    Steady-state  proton distributions with the trickle-down treatment~(dark-red solid line), the CEL approximation~(green dashed line) and the  neutron contribution to the trickle-down solution~(sky blue line) as functions of the nucleon energy. \textbf{Left}: scenario I.  \textbf{Right}: scenario II.
    }
    \label{fig:TD_vs_continuous_scenario_I}
\end{figure*}

Figure 4 compares the stationary nucleon distributions for the two scenarios. 
The difference between the TD and CEL treatment is visually quite striking. The latter shows a pronounced pile-up effect close to the maximum energy, with the number density of low-energy protons close to injection suppressed by almost two orders of magnitude compared to the TD approach.
In other words, in the physical TD approach,  we have to inject almost two orders of magnitude more protons compared to the artificial CEL case, although both solutions are normalized to the same energy density.  The large effect stems from the long confinement time of charged particles. In the CEL treatment, protons can undergo repeated stochastic acceleration and accumulate near the energy where acceleration is balanced by cooling; ultimately they are removed by the comparatively slow diffusive escape. In the TD treatment, instead, protons are removed by their conversion to neutrons which immediately escape or by their conversion to protons at lower energies;
 around $E^{*}_p\sim 10^4\,\mathrm{GeV}$, where $\tau_{\rm acc} \simeq \tau^{\rm CEL}_{cool, p}$, this creates an effective escape channel over a timescale of minutes to hours.

Therefore, the ratio between the non-thermal proton number densities normalizations in the TD and the CEL cases is approximately

\begin{equation}
    \frac{n^{\rm TD}_p  }{n^{\rm CEL}_p} \approx \tau_{\rm esc,p} (E^{*}_p) \Gamma^{\rm int}_p (E^{*}_p).
\end{equation}

This effect can be qualitatively understood with a very simple argument: for a fixed proton energy density $u_p$ and a fixed maximal proton energy $E_{p,\rm max}$, the total number density of protons $n_p\sim u_p/E_{p,\rm max}$ must come from the balance between the injection rate $q_p$ and the residence time of the protons $\tau_p$. Since the latter is substantially overestimated by the CEL approximation, the injection rate that fulfils the required energy density is vastly underestimated in that approach.

In addition, we also find as expected a strong change in the maximum proton energy in scenario II. The additional interaction channel, artificially neglected in the CEL approximation, limits acceleration to lower energies, with protons reaching a cutoff already below $10^5\,\mathrm{GeV}$, more than a factor 3 below the cutoff retrieved in the CEL approximation. In scenario I, where the dominant energy losses are anyway from BH interactions, this latter effect is less important.

\subsection{Neutrino emission}

\begin{figure}[t]
    \centering
    \includegraphics[width=0.49\linewidth]
    {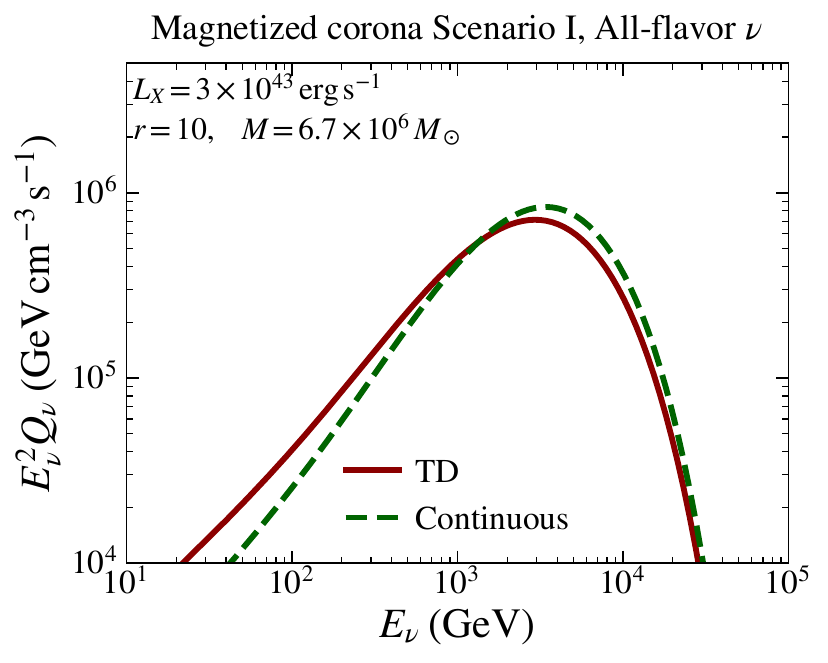}
    \includegraphics[width=0.49\linewidth]
    {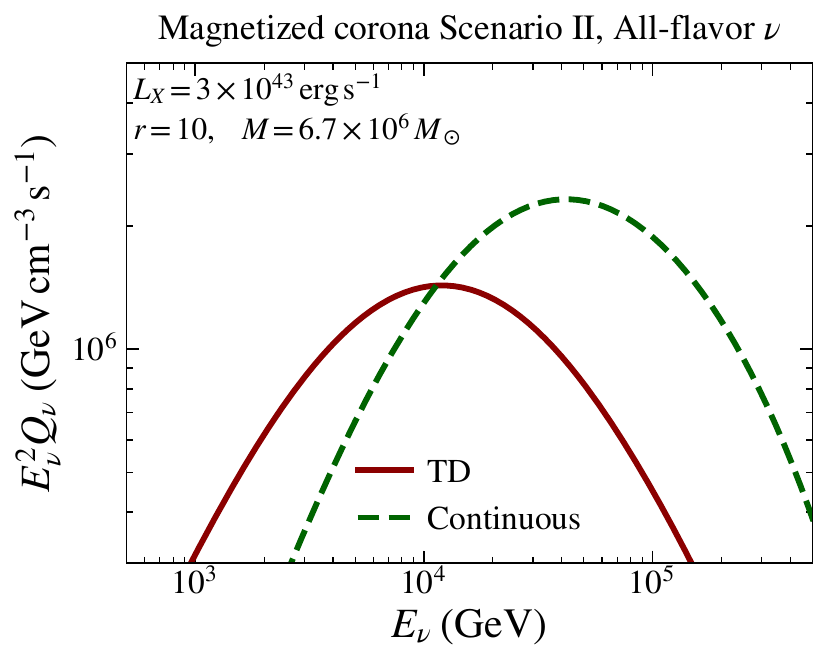}
    \caption{
    Neutrino production rates summed over all flavors, comparing the TD (solid dark-red lines) and CEL (dashed green lines) treatments. The left and right panels correspond to scenarios I
    and II, respectively.}
    \label{fig:neutrinos}
\end{figure}

These differences propagate into the all-flavor neutrino spectra shown in Fig.~\ref{fig:neutrinos}. Since the comparison is performed at fixed stationary proton energy density, the figure isolates changes in the energy distribution and interaction history of the nucleons: the much larger difference in the injection efficiency is basically invisible in the neutrino spectrum.

For scenario I (left panel), the two neutrino spectra remain relatively similar. The TD treatment modestly enhances the emission below the spectral maximum, while reducing the high-energy contribution and shifting the maximum to somewhat lower energy. The changes are much more pronounced in scenario II (right panel), as expected. The TD spectrum is enhanced at lower neutrino energies but declines much earlier at high energies; its maximum moves from several tens of TeV to less than 10~TeV, a shift by approximately a factor of four. For the benchmark parameters, the integrated neutrino luminosity is approximately 35\% lower than that obtained with the CEL approach, despite the identical normalization of the stationary proton energy density. As discussed in App.~\ref{app:secondary_cooling}, accounting for synchrotron cooling of secondary pions and muons reduces this difference to approximately 30\%, which we take as a representative estimate of the luminosity reduction when secondary cooling is included.

Thus, neutron escape has two distinct consequences in a turbulent corona. At fixed proton energy density, it modifies the neutrino spectrum, especially when photopion interactions limit acceleration. It also requires in both scenarios a much larger injection rate of protons to sustain a comparable neutrino emission, due to the inefficient confinement of the protons converted to neutrons. This sensitivity to confinement provides a useful contrast with the magnetic-reconnection scenario considered in the next section, where primary protons can already escape ballistically.

\section{Magnetic reconnection: numerical results}~\label{Sec:magnetic_reconnection}
\begin{figure*}[h!]
    \centering
    \includegraphics[width=0.48\textwidth]
    {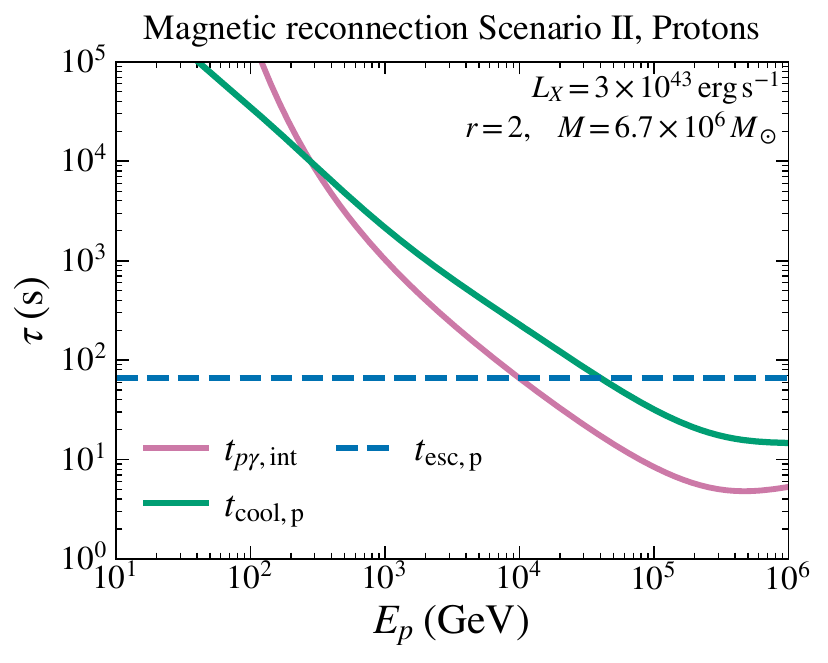}
    \hfill
    \includegraphics[width=0.48\textwidth]
    {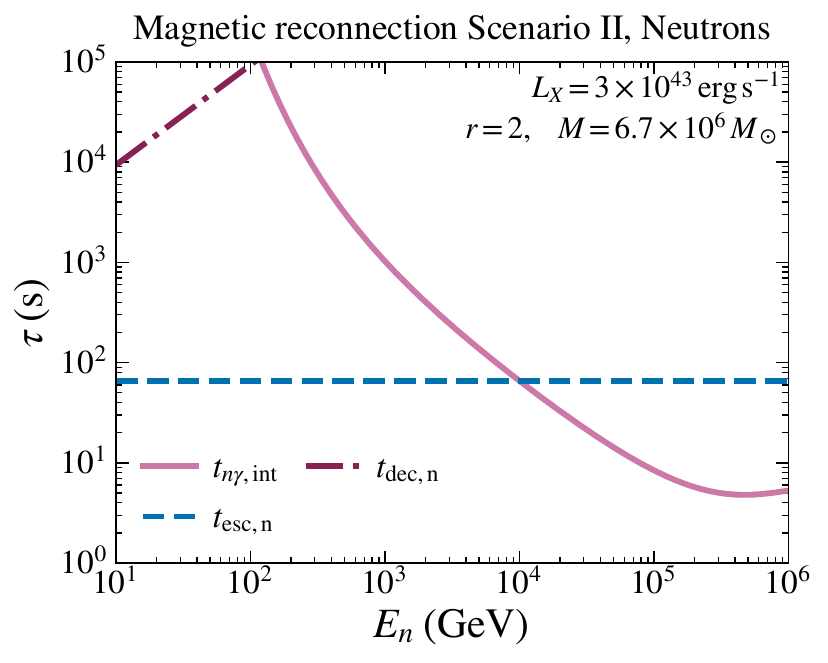}
    \caption{ Characteristic timescales for the magnetic-reconnection
    model with the background photon field of scenario II.
    \textbf{Left}: proton cooling, photopion interaction
    and escape timescales.
    \textbf{Right}: neutron photopion interaction, escape and decay timescales.
    }
    \label{fig:timescales_scenario_II_MR}
\end{figure*}

In this section, we discuss the results for the magnetic reconnection model. Fig.~\ref{fig:timescales_scenario_II_MR} shows the characteristic proton and neutron timescales.

Neutron escape is considerably faster than neutron decay, as in the magnetized turbulent-corona model. This effect is enhanced by the compactness of the region, which further reduces the crossing timescale. Moreover, the absence of the OUV photon component increases the relative importance of photopion interactions compared with Bethe--Heitler cooling, as already discussed for scenario II of the turbulent-corona model. However, the impact of neutron escape on the steady-state nucleon distribution is substantially reduced in the magnetic-reconnection model. Unlike in the turbulent corona, where repeated stochastic acceleration and energy-dependent confinement can lead to a pronounced high-energy pile-up, here the accelerated protons are injected with a prescribed broken power-law spectrum and escape on a timescale comparable to the light-crossing time of the interaction region. Therefore, although photohadronic collisions  provide an additional channel for removing protons from their original energy bins,  ballistic proton escape already limits the residence time of the accelerated particles. 
\begin{figure}[h!]
    \centering
    \includegraphics[width=0.49\linewidth]
    {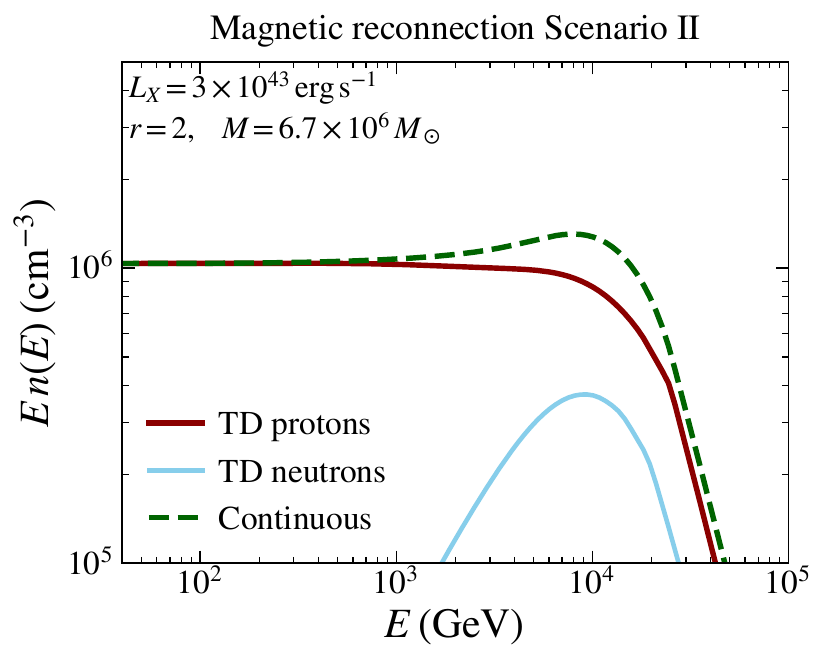}
    \includegraphics[width=0.49\linewidth]
    {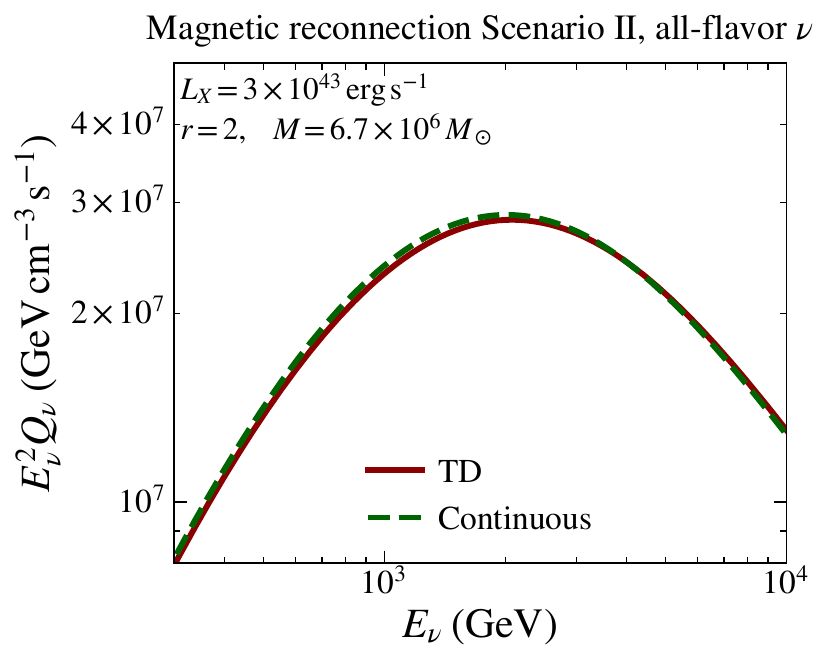}
    \caption{
    Results for the magnetic-reconnection model.
    \textbf{Left}:  Steady-state  proton distributions with the TD treatment~(dark-red solid line), the CEL approximation~(green dashed line) and the  neutron contribution to the trickle-down solution~(sky blue line) as functions of the nucleon energy. 
    \textbf{Right}: all-flavor neutrino production
    rates calculated using the normalized nucleon
    distributions for the TD (dark-red solid line)
    and CEL (green dashed line) treatments.
    }
    \label{fig:results_magnetic_reconnection}
\end{figure} The resulting nucleon distributions are shown in the left panel of Fig.~\ref{fig:results_magnetic_reconnection}.
In contrast to the magnetized turbulent-corona scenario, the CEL and TD treatments predict similar spectral shapes, with the peak of the total nucleon number density per logarithmic energy interval, $E n(E)$, shifted towards lower energies by approximately $20\%$ in the TD case. Nevertheless, the two treatments predict
different nucleon compositions: while the CEL approximation considers only protons, neutrons account for approximately $25\%$ of the total nucleon energy density
in the TD solution. Despite this substantial neutron contribution, the total nucleon energy densities differ
by only about $1.6\%$ between the two treatments. These differences become even less pronounced in the
neutrino sector, where both approaches predict nearly identical all-flavor neutrino production spectra,
as illustrated in the right panel of Fig.~\ref{fig:results_magnetic_reconnection}.

\section{Discussion and conclusions}
\label{Sec:conclusions}

Neutrons are a crucial degree of freedom in non-thermal environments. While they do not directly participate in particle energization in time-dependent magnetic fields, they take part in the dynamics through hadronic and photohadronic interactions that convert protons into neutrons. This dynamics is well known in several cosmic accelerators: we have highlighted its role in the coronae of AGN, a prominent candidate source for the emission of TeV neutrinos. Here, the simultaneous modeling of proton energization through interactions with magnetic turbulence and proton losses through hadronic and photohadronic interactions, has often led to modeling the latter as a CEL process. We go beyond this approximation and show that surprisingly large effects emerge.


On the one hand, modeling only the CEL part of the photohadronic interactions overestimates the characteristic energy-bin-removal timescale, because it only accounts for the average energy-loss frequency. Since the maximum energy that protons reach is determined by the competition between acceleration and interaction, the CEL treatment generally overestimates the maximum proton energy. On the other hand, the entity of this effect depends on whether photohadronic interactions are the dominant energy losses to begin with. For example, in our scenario I relating to the magnetized turbulent corona, the stronger BH losses are anyway dominant, and they are well-modeled by a CEL treatment, so the reduction in maximal energy is more modest.

A stronger effect comes instead from the proton-to-neutron conversion offering an escape route for protons that would otherwise be diffusively confined in the corona. By converting to neutrons, protons can escape from the corona within minutes to hours, rather than several days as expected from pure magnetic diffusion. As a result, a much larger injection rate of protons is required to sustain the stationary population producing high-energy neutrinos at the level suggested by the NGC~1068 signal. The enhancement is about two orders of magnitude. Of course, if protons already possess a faster escape route than magnetized diffusion, e.g.\ hydrodynamical turbulent diffusion~\cite{Lemoine:2024roa}, the corresponding enhancement would be more modest in proportion to this shorter escape timescale. This also explains why the effects of proton-to-neutron conversion are much smaller in our magnetic-reconnection scenario, where protons already escape rapidly from the corona.

The effect on nucleons is directly reflected in the neutrino emission. Compared with the CEL approach, the full treatment of photohadronic interactions shifts the neutrino spectrum towards lower energies, changes its spectral shape, and can reduce the neutrino luminosity by up to about 30\% in our benchmark models. Such differences are relevant for interpreting the neutrino emission of individual Seyfert galaxies~\cite{Carpio:2026xkf,Testagrossa:2026jcs}, and for predicting the diffuse neutrino flux from their population~\cite{Murase:2019vdl,Ambrosone:2024zrf,Fiorillo:2025ehn,Karavola:2026rpg}. In particular, the characteristic neutrino energy and spectral shape are not determined by the acceleration mechanism alone, but by its competition with proton confinement, charge-exchange interactions, and the different cooling channels. Neutrino searches based on physical spectral templates, rather than simple power laws, could help disentangle these effects.

The fate of the escaping neutrons is also interesting in another respect, in that they may lead to additional neutrinos, either from their direct beta decay, or from their hadronic or photohadronic interactions within the regions outside the corona. We do not investigate this question, which depends on the outer regions beyond the one-zone coronal scenario.

Our results are motivated by the NGC~1068 case. The quantitative impact of neutron escape will depend on the properties of individual sources, notably their photon fields, magnetic fields, and particle confinement timescales. Nevertheless, our work establishes a general point: in environments where particles are magnetically confined and accelerated, photohadronic interactions cannot always be reduced to a continuous energy loss. The conversion of energetic charged particles into freely escaping neutral ones can fundamentally change the energy budget of the accelerator, and thus the physical requirements for explaining its neutrino emission.

\textbf{Note added in preparation.} While this work was being finalized, a recent work appeared~\cite{Yuan:2026fte} also investigating particle acceleration and nuclear transport in Seyfert coronae, accounting for proton-to-neutron conversion as a catastrophic loss channel during acceleration. In that approach, secondary nucleons are evolved in a separate nuclear cascade, disentangled from acceleration. Instead, our treatment couples proton and neutron transport directly to stochastic acceleration, including the reinjection and subsequent acceleration of protons produced in hadronic and photohadronic interactions. While Ref.~\cite{Yuan:2026fte} focuses on the impact of nuclear composition on multimessenger emission, our work investigates the impact of discrete nucleon interactions and neutron escape on the stationary proton population and the resulting neutrino emission.

\begin{acknowledgments}

AA is thankful to Fabio Convenga for sharing useful suggestions on the graphical style and layout of the figures presented in this work.
AA gratefully acknowledges the Pierre Auger  Collaboration for helpful discussions that contributed to this work. DFGF is grateful to Luca Comisso, Maria Petropoulou, Enrico Peretti, and Federico Testagrossa for helpful discussions on this work.  AA and DFGF acknowledge the support by the research project TAsP (Theoretical Astroparticle Physics) funded by the Istituto Nazionale di Fisica Nucleare (INFN).

\end{acknowledgments}

\appendix

\section{Details of the photopion treatment}
\label{sec:app_photopion}

In this appendix, we specify how the microscopic photopion properties are extracted from event simulations and converted into the coefficients entering the CEL  and TD  prescriptions. In particular, we focus on the microscopic channel identification, the photon-field averages, and the specific assumptions underlying the implementation.

\subsection{Microscopic photopion properties}
\label{app:microscopic_photopion}

We use Chromo~\cite{Fedynitch:2025mgj} with the SOPHIA 2.0 event generator~\cite{Mucke:1999yb} to determine the inelastic photopion cross section $\sigma_{p\gamma}(\varepsilon')$ and the properties of the final-state baryons. Photon--proton collisions are simulated for different photon energies $\varepsilon'$ in the proton rest frame. Following Refs.~\cite{Collins:1997nm,Mucke:1999yb}, we assign to each final-state particle $i$ the light-cone energy fraction
\begin{equation}
\xi_i^-\equiv\frac{E'_i-p'_{z,i}c}{m_pc^2},
\label{eq:app_lightcone_fraction}
\end{equation}
where the incident photon defines the positive longitudinal direction. In the ultrarelativistic limit, $\xi_i^-$ approximates the fraction of the incident proton energy carried by particle $i$ in the coronal frame.
For TD transport, we select the final-state proton or neutron with the largest $\xi_i^-$ as the \emph{leading nucleon}. 
The microscopic probability for an incident proton to produce a leading nucleon of species $b\in\{p,n\}$ and its conditional mean energy fraction are
\begin{equation}
p_{p\rightarrow b}(\varepsilon')=\frac{N_{p\rightarrow b}(\varepsilon')}{N_{\rm ev}(\varepsilon')},\qquad\xi_{p\rightarrow b}^{\rm MC}(\varepsilon')=\left\langle\xi_b^-\right\rangle_{p\rightarrow b},
\label{eq:app_microscopic_probability}
\end{equation}
where $N_{\rm ev}$ is the number of simulated inelastic interactions at a given $\varepsilon'$ and $N_{p\rightarrow b}$ counts events assigned to channel $p\rightarrow b$. The average is taken only over events assigned to that channel. The corresponding microscopic channel inelasticity, $\kappa_{p\rightarrow b}^{\rm MC}(\varepsilon')=1-\xi_{p\rightarrow b}^{\rm MC}(\varepsilon')$, describes the mean energy not retained by that leading nucleon.

For the CEL, we define 
\begin{equation}
\kappa_B^{\rm MC}(\varepsilon')\equiv1-\left\langle\xi_B^-\right\rangle,
\label{eq:app_baryon_inelasticity}
\end{equation}
with the average taken over events for which a \emph{leading baryon} is identified. In other words, we do not restrict the leading particle to be a proton or neutron, allowing for other baryonic species in the final state. This choice is motivated by the CEL approximation, in which inelastic interactions are described through their average energy losses, while the proton identity is assumed to remain unchanged throughout the transport.
This quantity is extracted directly from the Monte Carlo events, rather than reconstructed from the two propagated leading-nucleon channels. Consequently, the microscopic inputs used for CEL and TD treatments do not coincide exactly.

\begin{figure}[t]
    \centering
    \includegraphics[width=0.49\linewidth]{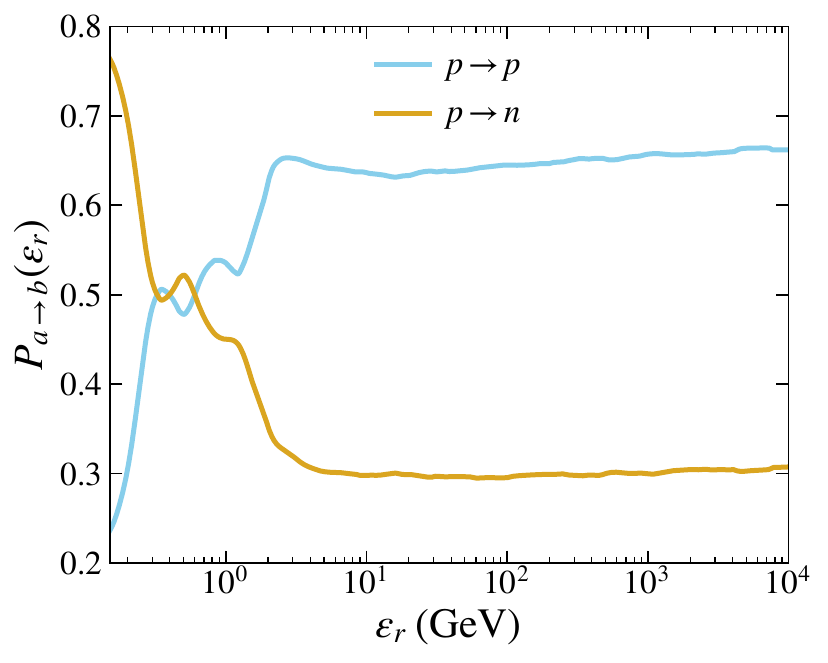}
    \includegraphics[width=0.49\linewidth]{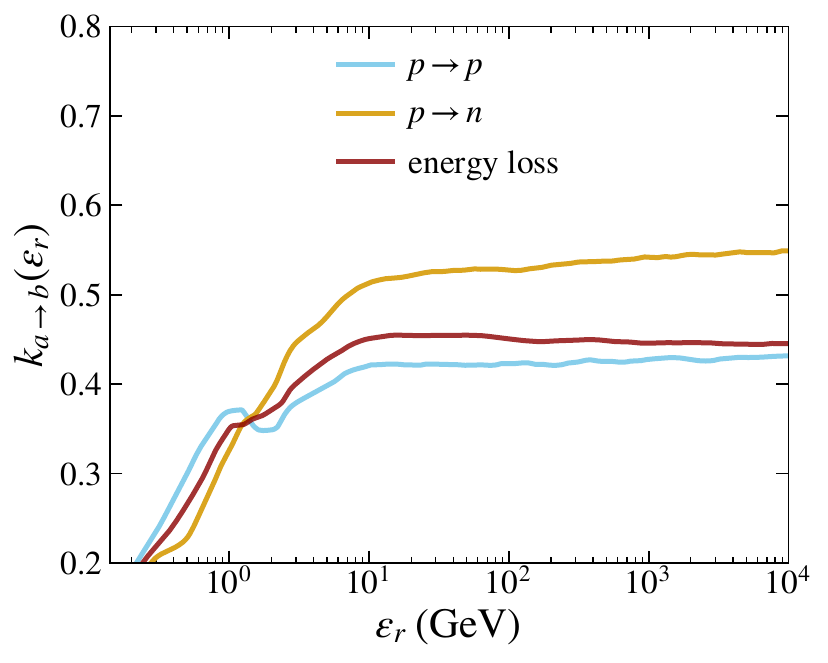}
    \caption{Microscopic photopion properties as functions of the incident photon energy $\varepsilon'$ in the proton rest frame. \textbf{Left}: probabilities $p_{p\rightarrow p}$ and $p_{p\rightarrow n}$ for the leading-nucleon channels. \textbf{Right}: corresponding channel inelasticities $\kappa_{p\rightarrow p}^{\rm MC}$ and $\kappa_{p\rightarrow n}^{\rm MC}$, compared with the leading-baryon inelasticity $\kappa_B^{\rm MC}$ employed for CEL cooling.}
    \label{fig:details_photopion}
\end{figure}

As shown in Fig.~\ref{fig:details_photopion}, both the channel probabilities and energy fractions depend on $\varepsilon'$, particularly near the photopion threshold. At higher photon energies, $p_{p\rightarrow p}$ and $p_{p\rightarrow n}$ approach approximately $2/3$ and $1/3$, respectively. The channel inelasticities are typically of order $20\%$ near threshold and increase with $\varepsilon'$. The leading-baryon quantity $\kappa_B^{\rm MC}$ is not, in general, identical to either channel inelasticity or to their probability-weighted average, since it is defined using a different final-state selection.

\subsection{Effective coefficients for CEL and TD transport}
\label{app:effective_photopion}

The microscopic coefficients determined in the previous subsection must be averaged over the radiation field of the coronal configuration, in order to determine the corresponding macroscopic quantities. Using the photon distribution and the function $J$ defined in Eq.~\ref{eq:td_photon_integral}, we introduce the differential interaction weight
\begin{equation}
\mathcal{W}_a(E,\varepsilon')\equiv\frac{c\,\sigma_{a\gamma}(\varepsilon')\varepsilon'}{2\gamma_a^2}J\!\left(\frac{\varepsilon'}{2\gamma_a}\right),
\label{eq:app_photopion_weight}
\end{equation}
where $a\in\{p,n\}$. This weight has dimensions of an interaction rate per unit $\varepsilon'$; its integral over $\varepsilon'\geq\varepsilon'_{\rm th}$ gives the total rate already defined in Eq.~\eqref{eq:photopion_interaction_rate}. All photon energies and tabulated cross sections are evaluated in consistent units.

For TD transport, weighting the interaction probability by the microscopic channel probability yields the partial interaction rate (directly evaluated for $a=p$; the neutron channels are constructed as described below)
\begin{equation}
\Gamma_{a\rightarrow b}^{(a\gamma)}(E)=\int_{\varepsilon'_{\rm th}}^\infty d\varepsilon'\,\mathcal{W}_a(E,\varepsilon')\,p_{a\rightarrow b}(\varepsilon'),
\label{eq:app_channel_rate}
\end{equation}
and the channel-conditioned mean daughter energy fraction is
\begin{equation}
\xi_{a\rightarrow b}^{(a\gamma)}(E)=\frac{\int_{\varepsilon'_{\rm th}}^\infty d\varepsilon'\,\mathcal{W}_a(E,\varepsilon')\,p_{a\rightarrow b}(\varepsilon')\,\xi_{a\rightarrow b}^{\rm MC}(\varepsilon')}{\Gamma_{a\rightarrow b}^{(a\gamma)}(E)}.
\label{eq:app_effective_xi}
\end{equation}
The corresponding channel inelasticity is $k_{a\rightarrow b}^{(a\gamma)}(E)=1-\xi_{a\rightarrow b}^{(a\gamma)}(E)$, consistently with Eq.~\eqref{eq:td_energy_mapping}. The rate and the energy fraction above are inserted into the general reinjection formula, Eq.~\eqref{eq:td_injection}. In particular, the Jacobian contains the derivative of the \emph{photon-field-averaged} function $\xi_{a\rightarrow b}^{(a\gamma)}(E)$, not the derivative of the microscopic Monte Carlo table.

The effective photopion inelasticity entering CEL cooling is computed separately from the leading-baryon observable,
\begin{equation}
K_{p\gamma,\rm loss}(E)=\frac{\int_{\varepsilon'_{\rm th}}^\infty d\varepsilon'\,\mathcal{W}_p(E,\varepsilon')\,\kappa_B^{\rm MC}(\varepsilon')}{\Gamma_{p\gamma}(E)}.
\label{eq:app_effective_loss_k}
\end{equation}
This coefficient enters the CEL loss time through Eq.~\eqref{eq:tloss_photopion}. It is important to distinguish $K_{p\gamma,\rm loss}$ from the TD channel inelasticities $k_{p\rightarrow b}^{(p\gamma)}$: the former follows the energy of the leading baryon irrespective of species, while the latter follow only explicitly propagated leading proton or neutron channels. Both prescriptions use the same underlying interaction simulations and photon-field weight, but they employ different microscopic observables.

For diagnostic purposes, we define the effective branching fractions
\begin{equation}
B_{a\rightarrow b}^{(a\gamma)}(E)\equiv\frac{\Gamma_{a\rightarrow b}^{(a\gamma)}(E)}{\Gamma_{a\gamma}(E)}.
\label{eq:app_effective_branching}
\end{equation}
They are not independent inputs when the partial photopion rates are calculated directly through Eq.~\eqref{eq:app_channel_rate}.

\begin{figure}[t]
    \centering
    \includegraphics[width=0.49\linewidth]{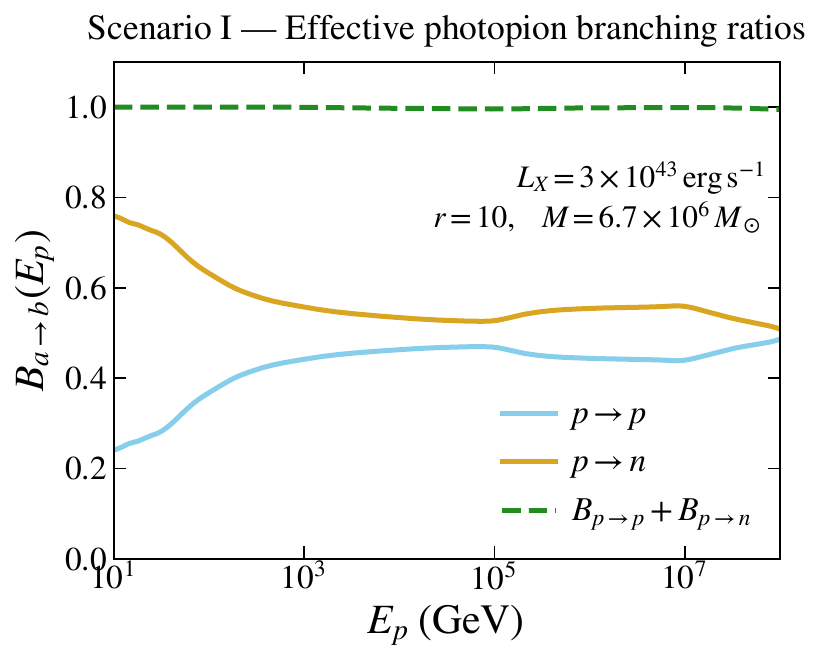}
    \includegraphics[width=0.49\linewidth]{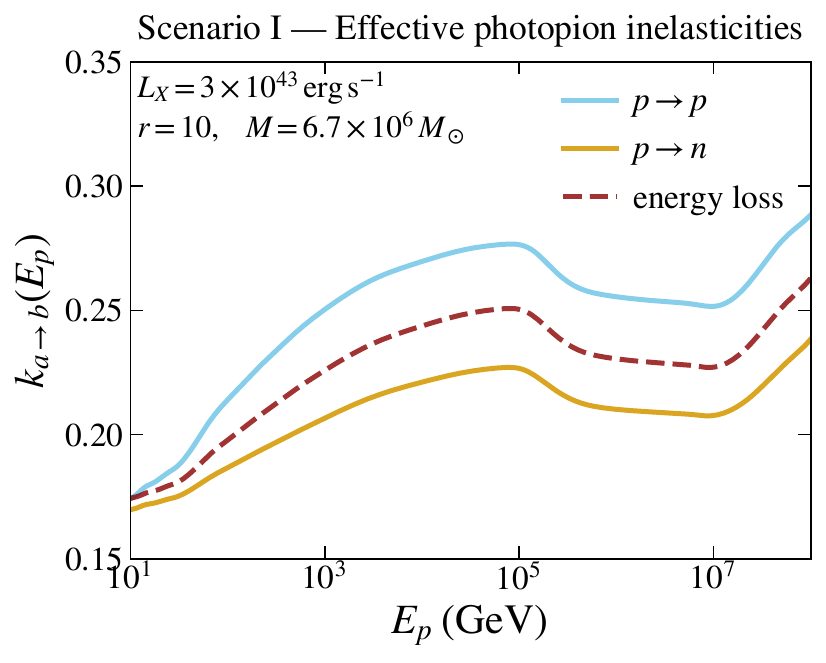}
    \caption{Photopion coefficients after convolution with the photon field of Scenario I, as functions of the parent proton energy. \textbf{Left}: effective branching fractions $B_{p\rightarrow p}^{(p\gamma)}$ and $B_{p\rightarrow n}^{(p\gamma)}$ and their sum. \textbf{Right}: TD channel inelasticities $k_{p\rightarrow p}^{(p\gamma)}$ and $k_{p\rightarrow n}^{(p\gamma)}$, compared with the CEL coefficient $K_{p\gamma,\rm loss}$.}
    \label{fig:details_photopion_averaged}
\end{figure}

Figure~\ref{fig:details_photopion_averaged} illustrates the effect of averaging over the target photon distribution in Scenario I. The two TD energy fractions are conditional on different final-state channels, whereas $K_{p\gamma,\rm loss}$ is averaged from the leading-baryon observable. These distinctions are preserved when constructing the transport coefficients.

\subsection{Unassigned channels and neutron-initiated interactions}
\label{app:photopion_approximations}

As said above, not every simulated photopion event can be assigned to a leading proton or neutron.  In this subsection, we investigate their impact over the TD equations. To this purpose, we define the  microscopic residual probability and its photon-field-averaged counterpart as,
\begin{equation}
\begin{aligned}f_{\rm noN}^{\rm MC}(\varepsilon')&=1-p_{p\rightarrow p}(\varepsilon')-p_{p\rightarrow n}(\varepsilon'),\\ B_{\rm noN}(E)&=1-\sum_{b\in\{p,n\}}B_{p\rightarrow b}^{(p\gamma)}(E).\end{aligned}
\label{eq:app_missing_branching}
\end{equation}
Such interactions are retained in the total interaction rate, but do not produce an explicit leading-proton or leading-neutron reinjection. The microscopic residual reaches approximately $7\%$ at some photon energies in the interpolated input tables. After averaging over the photon field of Scenario I, the residual remains below approximately $0.5\%$ in the energy range relevant for photopion transport. Its fraction among interactions sampled by the stationary TD proton population is
\begin{equation}
f_{\rm noN}^{\rm int}\equiv\frac{\int dE\,n_p^{\rm TD}(E)\,\Gamma_{p\gamma}(E)\,B_{\rm noN}(E)}{\int dE\,n_p^{\rm TD}(E)\,\Gamma_{p\gamma}(E)}.
\label{eq:app_missing_weighted}
\end{equation}
We obtain $f_{\rm noN}^{\rm int}\simeq0.33\%$ for Scenario I. Therefore, we conclude that the impact of un-assigned events cannot impact the solutions of the TD transport equations. 

The microscopic channel tables described above are obtained for photon--proton collisions. For neutron-initiated interactions, the total rate $\Gamma_{n\gamma}(E)$ is evaluated independently, while the effective branching fractions and daughter energy fractions are related to the proton channels by approximate isospin symmetry,
\begin{equation}
\begin{aligned}
B_{n\rightarrow n}^{(n\gamma)}(E)&\simeq B_{p\rightarrow p}^{(p\gamma)}(E), & B_{n\rightarrow p}^{(n\gamma)}(E)&\simeq B_{p\rightarrow n}^{(p\gamma)}(E),\\
\xi_{n\rightarrow n}^{(n\gamma)}(E)&\simeq\xi_{p\rightarrow p}^{(p\gamma)}(E), & \xi_{n\rightarrow p}^{(n\gamma)}(E)&\simeq\xi_{p\rightarrow n}^{(p\gamma)}(E).
\end{aligned}
\label{eq:app_isospin_approximation}
\end{equation}

Therefore, the neutron partial rates are constructed as $\Gamma_{n\rightarrow b}^{(n\gamma)}(E)=B_{n\rightarrow b}^{(n\gamma)}(E)\Gamma_{n\gamma}(E)$. The energy derivatives required by the TD Jacobian are inherited from the corresponding proton energy fractions. This approximation does not assume that $\Gamma_{n\gamma}=\Gamma_{p\gamma}$.

\section{Details of the hadronuclear treatment}
\label{sec:app_hadronuclear}

In this section, we describe the microscopic inputs and their use in the CEL and TD transport prescriptions for hadronuclear collisions obtained from Chromo~\cite{Fedynitch:2025mgj} with Sibyll~2.3e~\cite{Ahn:2009wx,Fedynitch:2018cbl,Riehn:2019jet}. 
In particular, we specify the cross sections, the leading-nucleon selection, and the effective coefficients entering those expressions.

\subsection{Inelastic hadronuclear cross sections}
\label{app:hadronic_cross_sections}

We extract the inelastic cross sections from the event generator as functions of the projectile energy. Since in the Chromo--Sibyll setup, the energy range starts at $E=60\,\mathrm{GeV}$~\cite{Riehn:2019jet}, we construct a hybrid cross section by combining the Monte Carlo predictions with the analytical parametrization of Ref.~\cite{Kelner:2006tc}. In particular, we adopt
\begin{equation}
\sigma_{pp}^{\rm inel}(E)=
\begin{cases}
\sigma_{pp}^{\rm KA}(E),
& E\leq E_{\rm match},
\\[3pt]
\sigma_{pp}^{\rm Sibyll}(E),
& E>E_{\rm match},
\end{cases}
\label{eq:app_sigma_pp_hybrid}
\end{equation}
where $E_{\rm match}=100\,\mathrm{GeV}$. In other words, we allow the analytical prescription to describe the lower energies while retaining the Monte Carlo predictions at higher energies.
The analytical cross section is given by~\cite{Kelner:2006tc}
\begin{equation}
\begin{aligned}
\sigma_{pp}^{\rm KA}(E)
={}&
\left[
34.3+1.88L+0.25L^2
\right]
\\
&\times
\left[
1-
\left(
\frac{E_{\rm th}}{E}
\right)^4
\right]^2
\,\mathrm{mb},
\end{aligned}
\label{eq:app_sigma_KA}
\end{equation}
where $L=\ln(E/\mathrm{TeV})$ and $E_{\rm th}=1.22\,\mathrm{GeV}$ is the kinematic energy threshold.
For neutron--proton interactions, we employ the same treatment simply assuming that $\sigma_{np}^{\rm inel}(E)\simeq \sigma_{pp}^{\rm inel}(E)$ for $E \le 100\, \rm GeV$.
\begin{figure}[t]
    \centering
    \includegraphics[width=0.92\linewidth]
    {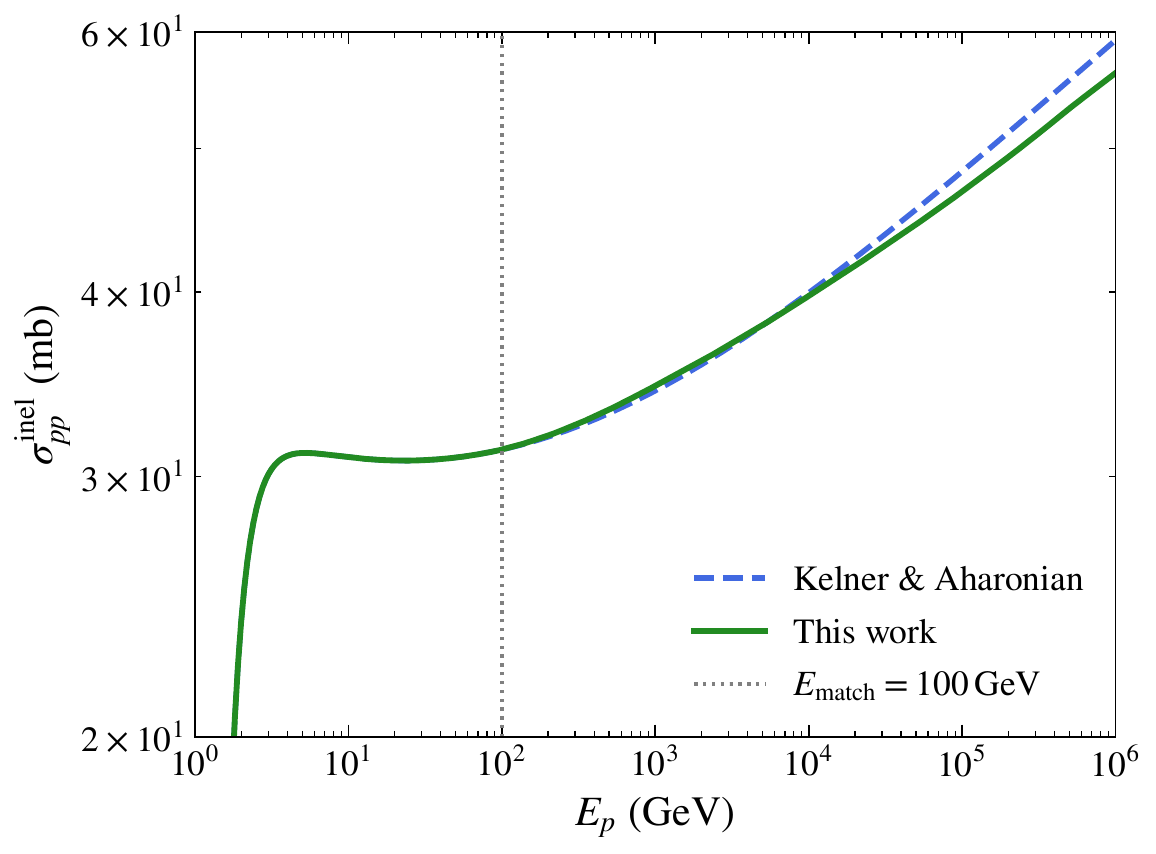}
    \caption{
    Inelastic hadronuclear cross sections
    as functions of the projectile energy for pp interactions.
    The function used in this work is represented by a solid green line while the analytical parametrization of Ref.~\cite{Kelner:2006tc} is shown as a dashed blue line.
    The vertical dotted line indicates  the matching energy of
    $E_{\rm match}=100\,\mathrm{GeV}$.
    }
    \label{fig:app_hadronic_cross_sections}
\end{figure} The adopted prescription, for pp, is shown in Fig.~\ref{fig:app_hadronic_cross_sections} for reference.

\subsection{Microscopic leading-nucleon properties}
\label{app:hadronic_leading_nucleons}

To determine the energy redistribution in each channel, we generate inelastic events for both proton and neutron projectiles. We identify the leading final-state nucleon among outgoing protons and neutrons with positive longitudinal momentum ($p_{z,i}>0$), selecting the one that maximizes the forward light-cone fraction~\cite{Riehn:2019jet}
\begin{equation}
x_i^+\equiv\frac{E_i+p_{z,i}c}{E+p_{\rm proj}c}.
\label{eq:app_hadronic_lightcone}
\end{equation}
where $E_i$ and $p_{z,i}$ are the energy and longitudinal momentum of the final-state nucleon, and $E$ and $p_{\rm proj}$ refer to the incident nucleon.
The fraction of the projectile energy retained by the selected nucleon is
\begin{equation}
\xi_i=\frac{E_i}{E}.
\label{eq:app_hadronic_energy_fraction}
\end{equation}
This quantity, rather than the light-cone fraction itself, is used to determine the energy of the daughter nucleon in the TD transport calculation.

The microscopic branching fraction and conditional mean daughter energy fraction are given by
\begin{equation}
B_{a\rightarrow b}^{(ap)}(E)=\frac{N_{a\rightarrow b}(E)}{N_{\rm ev}(E)},\qquad\xi_{a\rightarrow b}^{(ap)}(E)=\left\langle\frac{E_b}{E}\right\rangle_{a\rightarrow b}.
\label{eq:app_hadronic_branching}
\end{equation}
The denominator $N_{\rm ev}$ includes all simulated inelastic events, whereas the conditional average is restricted to events with a leading nucleon of species $b$. The channel inelasticity is $k_{a\rightarrow b}^{(ap)}(E)=1-\xi_{a\rightarrow b}^{(ap)}(E)$. 
As for photohadronic interactions, also in this case, a small fraction of events cannot be assigned to either leading-nucleon channel.
The corresponding residual probability is given by

\begin{equation}
B_{a,\rm res}^{(ap)}(E)=1-\sum_{b\in\{p,n\}}B_{a\rightarrow b}^{(ap)}(E).
\label{eq:app_hadronic_residual}
\end{equation}
Such events are included in the total interaction rate but not in either reinjection channel.

The effective inelasticity associated with the retained energy of the explicitly
propagated leading nucleons is defined as
\begin{equation}
K_{ap}^{\rm eff}(E)
=
1-
\sum_{b=p,n}
B_{a\rightarrow b}^{(ap)}(E)
\xi_{a\rightarrow b}^{(ap)}(E).
\label{eq:app_hadronic_k_eff}
\end{equation}
In this definition, interactions without an identified leading proton or neutron contribute to the total interaction rate but do not produce
explicit nucleon reinjection terms. Consequently, any energy retained by baryonic secondaries in these residual channels is not tracked in the
nucleon transport calculation. We find this approximation reasonable given that hadronuclear processes are subdominant to shape the overall cosmic rays dynamics in the magnetized turbulent corona scenario~(see Fig.~\ref{fig:timescales_scenario_I}). 

\begin{figure}[t]
    \centering
    \includegraphics[width=0.49\linewidth]{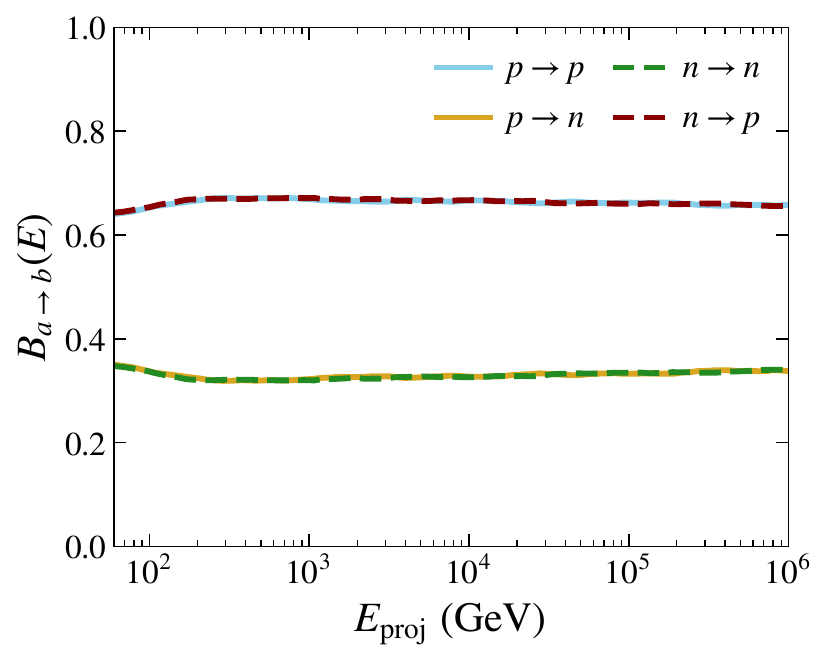}
    \includegraphics[width=0.49\linewidth]{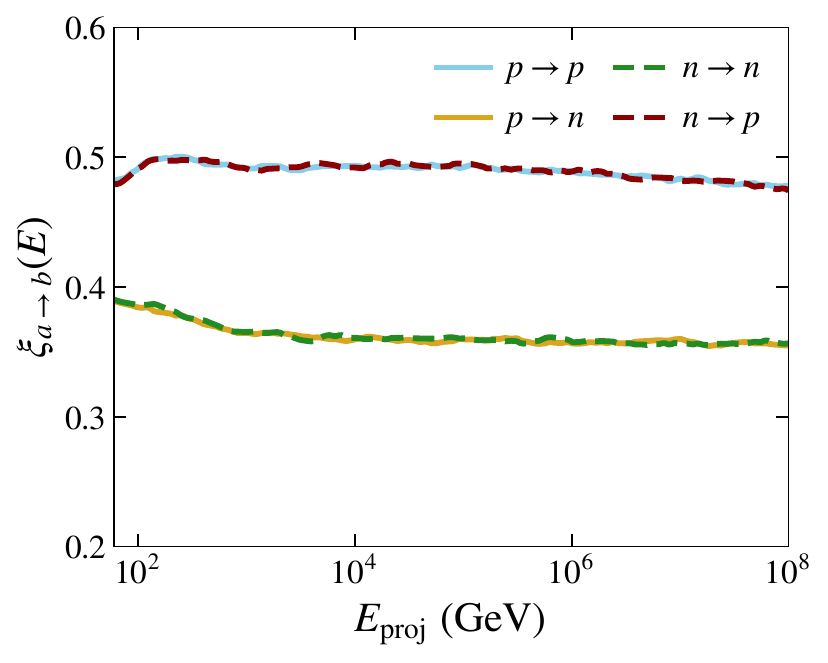}
    \caption{Leading-nucleon properties of inelastic $pp$ and $np$ collisions obtained with Chromo and Sibyll 2.3e. \textbf{Left}: branching fractions for the $p\rightarrow p$, $p\rightarrow n$, $n\rightarrow n$, and $n\rightarrow p$ channels. \textbf{Right}: corresponding mean daughter energy fractions $\xi_{a\rightarrow b}^{(ap)}$.}
    \label{fig:app_hadronic_effective}
\end{figure}

 Fig.~\ref{fig:app_hadronic_effective} shows that the branching fractions and mean daughter energy fractions exhibit relatively weak energy dependence across the range relevant for the present calculation. In particular, the effective leading-nucleon inelasticities constructed from the channel coefficients vary by less than approximately $1\%$ in the range considered. Therefore, we evaluate $B_{a\rightarrow b}^{(ap)}$ and $\xi_{a\rightarrow b}^{(ap)}$ at $E_{\rm ref}=10\,\mathrm{TeV}$ and keep them constant in the transport calculation. The Monte Carlo energy dependence is smoothed using a five-point moving average and interpolated in logarithmic projectile energy before extracting the reference values. The resulting coefficients are listed in Table~\ref{tab:app_hadronic_parameters}.

\begin{table}[t]
    \centering
    \caption{Effective hadronuclear parameters at $E_{\rm ref}=10\,\mathrm{TeV}$ obtained from Sibyll 2.3e. In the $pp$ column the initial nucleon is $a=p$; in the $np$ column it is $a=n$. The effective inelasticity $K_{ap}^{\rm eff}$ is constructed from the same leading-nucleon coefficients used in the TD reinjection.}
    \label{tab:app_hadronic_parameters}
    \renewcommand{\arraystretch}{1.15}
    \begin{tabular}{lcc}
        \hline
        Quantity & $pp$ & $np$ \\
        \hline
        $B_{a\rightarrow p}^{(ap)}$ & 0.6665 & 0.6673 \\
        $B_{a\rightarrow n}^{(ap)}$ & 0.3272 & 0.3261 \\
        $\xi_{a\rightarrow p}^{(ap)}$ & 0.4934 & 0.4920 \\
        $\xi_{a\rightarrow n}^{(ap)}$ & 0.3595 & 0.3607 \\
        $K_{ap}^{\rm eff}$ & 0.5536 & 0.5540 \\
        \hline
    \end{tabular}
\end{table}

Equipped with all these quantities, analogously to the photohadronic collisions, we define the  partial interaction rates entering
the TD reinjection terms  by
\begin{equation}
\Gamma_{a\rightarrow b}^{(ap)}(E)
=
B_{a\rightarrow b}^{(ap)}(E_{\rm ref})
\Gamma_{ap}(E),
\label{eq:app_hadronic_partial_rates}
\end{equation}
where $\Gamma_{ap}$ is obtained from the cross section of Eq.~\ref{eq:app_sigma_pp_hybrid}. Likewise, the energy mapping between
parent and daughter nucleons becomes
\begin{equation}
E=
\xi_{a\rightarrow b}^{(ap)}(E_{\rm ref})E_*.
\label{eq:app_hadronic_energy_mapping}
\end{equation}
Consequently, the TD reinjection term
takes the simplified form
\begin{equation}
Q_{b\leftarrow a}^{(ap)}(E)
=
\frac{
\Gamma_{a\rightarrow b}^{(ap)}(E_*)
n_a(E_*)
}{
\xi_{a\rightarrow b}^{(ap)}(E_{\rm ref})
},
\label{eq:app_hadronic_reinjection}
\end{equation}
where
$E_*=E/\xi_{a\rightarrow b}^{(ap)}(E_{\rm ref})$.

\section{Numerical implementation of the transport equations}
\label{app:numerical_transport}

In this appendix, we describe the numerical procedures adopted to solve the steady-state transport equations presented in the main text. Both acceleration scenarios are evaluated on a logarithmically spaced energy grid covering $E_{\min}=10\,\mathrm{GeV}$ to $E_{\max}=10^8\,\mathrm{GeV}$, with 220 grid points. The numerical treatment differs between the magnetized turbulent-corona and magnetic-reconnection scenarios because stochastic acceleration is explicitly included only in the former.

\subsection{Fokker--Planck discretization}

In the magnetized turbulent-corona scenario, the proton transport equation contains both stochastic acceleration and continuous energy losses. We solve the resulting second-order differential equation using a finite-volume discretization in energy space.

The differential terms of the transport equation are expressed through the energy-space particle flux. We denote by $E_i$ the grid energies and define the interfaces between adjacent cells as $E_{i+1/2}=\sqrt{E_iE_{i+1}}$. The corresponding cell width is $\Delta E_i=E_{i+1/2}-E_{i-1/2}$, with the outer interfaces obtained by linear extrapolation from the adjacent grid points.

The stochastic-acceleration contribution to the flux at the interface between cells $i$ and $i+1$ is discretized as
\begin{equation}
\mathcal{F}_{i+1/2}^{\rm acc}=-\frac{A_{i+1/2}}{E_{i+1}-E_i}\left(\frac{n_{i+1}}{E_{i+1}^2}-\frac{n_i}{E_i^2}\right),
\label{eq:app_acc_flux}
\end{equation}
where $A_{i+1/2}=\sqrt{A_iA_{i+1}}$ and $A_i=E_i^4/\tau_{\rm acc}(E_i)$.

Since continuous energy losses transport particles from higher to lower energies, we adopt a first-order upwind prescription for the corresponding flux,
\begin{equation}
\mathcal{F}_{i+1/2}^{\rm loss}=-b_{i+1/2}n_{i+1},
\label{eq:app_loss_flux}
\end{equation}
where $b_{i+1/2}=\sqrt{b_i b_{i+1}}$. The total energy-space flux is obtained by summing the two contributions.

The steady-state transport equation is then discretized in conservative form,
\begin{equation}
\frac{\mathcal{F}_{i+1/2}-\mathcal{F}_{i-1/2}}{\Delta E_i}+s_i n_i=Q_i,
\label{eq:app_fv_equation}
\end{equation}
where $s_i$ collects the escape and, when applicable, the discrete interaction rates, while $Q_i$ contains the primary injection and secondary reinjection terms.

The resulting linear system is solved by imposing a vanishing energy-space flux at the lower boundary and an absorbing condition at the upper boundary~\cite{Yuan:2025ctq},
\begin{equation}
\mathcal{F}_E(E_{\min})=0,\qquad n_p(E_{\max})=0.
\label{eq:app_fp_boundary}
\end{equation}
The first condition is implemented by excluding any flux through the lower external cell interface, while the second is imposed directly on the last row of the discretized operator.

\subsection{Magnetic-reconnection discretization}

In the magnetic-reconnection scenario, proton acceleration is incorporated into the prescribed injection spectrum rather than explicitly modeled in the radiative-region transport equation. Consequently, the equation contains only continuous energy losses, escape, and, in the TD approach, discrete interaction sinks and reinjection terms.

The first-order energy derivative is discretized using an upwind scheme. Denoting by $b_i$ the appropriate continuous energy-loss rate and by $s_i$ the sum of the escape and discrete interaction rates, the resulting equation reads
\begin{equation}
\frac{b_i n_i-b_{i+1}n_{i+1}}{E_{i+1}-E_i}+s_i n_i=Q_i.
\label{eq:app_mr_discretization}
\end{equation}
For the CEL calculation, $s_i$ contains only proton escape, whereas in the TD calculation it also includes the total photohadronic interaction rate.
The discretized operator is upper bidiagonal and can therefore be solved by backward substitution, starting from the absorbing boundary $n_p(E_{\max})=0$. No additional lower-energy boundary condition is required.

\subsection{Iterative solution and convergence for the TD equations}

The TD treatment requires the numerical evaluation of the secondary source terms associated with the different nucleon production channels. For each channel $a\rightarrow b$, the parent-to-daughter energy mapping defined in Eq.~\eqref{eq:td_energy_mapping} is evaluated on the energy grid. The resulting relation between parent and daughter energies is numerically inverted by interpolation to determine the parent energy $E_*$ corresponding to each daughter energy $E$.

The partial interaction rate, parent particle distribution, and effective daughter energy fraction are then interpolated at $E_*$. The Jacobian entering Eq.~\eqref{eq:td_injection} is evaluated using the energy derivative of the effective daughter energy fraction. For photohadronic interactions, these derivatives are computed numerically from the energy-dependent effective fractions using second-order accurate finite differences. For hadronuclear interactions, the adopted constant energy fractions imply a vanishing derivative, reducing the Jacobian to the corresponding daughter energy fraction.

The coupled proton--neutron TD equations are solved using a fixed-point iterative procedure. Starting from an initial estimate of the particle distributions, we evaluate the photohadronic and hadronuclear reinjection terms and update the proton and neutron solutions. The updated distributions are then used to reconstruct the secondary source terms at the following iteration.

To improve numerical stability, the particle distributions are updated through under-relaxation,
\begin{equation}
n_a^{(k+1)}(E)=(1-\alpha)n_a^{(k)}(E)+\alpha n_a^{\rm raw}(E),\qquad a\in\{p,n\},
\label{eq:app_under_relaxation}
\end{equation}
where $n_a^{\rm raw}$ denotes the solution obtained before mixing and $\alpha$ is the relaxation parameter.

For the magnetized turbulent-corona scenario, we initialize the proton distribution using the solution without secondary reinjection and set the initial neutron distribution to zero. We adopt $\alpha=0.2$, a relative convergence tolerance of $10^{-8}$, and a maximum of 2000 iterations. Convergence is monitored through the maximum relative change of both nucleon distributions, with a small numerical floor introduced to avoid artificially large relative differences in regions where the distributions are negligible.

For magnetic reconnection, both particle distributions are initially set to zero. We employ the same relaxation parameter, $\alpha=0.2$, with a relative tolerance of $10^{-9}$ and a maximum of 1000 iterations. In this case, the convergence criterion is evaluated from the changes between the current distributions and the corresponding unmixed updates.

After the linear solves, numerical safeguards are applied to remove non-finite entries and small negative values from the particle distributions.

\section{Impact of secondary pion and muon cooling}
\label{app:secondary_cooling}

The neutrino production prescription adopted in the main text assumes that unstable secondaries decay without significant energy losses. In this appendix, we quantitatively assess this approximation  by explicitly following the production, synchrotron cooling,
and decay of charged pions and muons.
We use the same parameters and the steady-state solutions as obtained in the main text for the CEL and TD treatments. We start by discussing the pions and muons timescales, then we report their transport equation and finally how we evaluate the neutrino emissions from their decay.

\subsection{Synchrotron cooling and decay timescales}
\label{app:secondary_timescales}

For a relativistic charged particle $a={\pi^\pm,\mu^\pm}$, the decay timescale scales linearly with the energy as for the decay timescale of neutrons. By contrast, the synchrotron cooling timescale decreases with energy and magnetic field strength.  The critical energy at which the two timescales
are equal is given by 
\begin{equation}
  E_{a,\mathrm{crit}}=
  \left(\frac{6\pi m_a^5 c^5}
  {\sigma_T m_e^2 B^2\tau_{a,0}}\right)^{1/2},
  \label{eq:app_sec_critical_energy}
\end{equation}
where  $m_a$ and $\tau_{a,0}$ are the mass and proper lifetime of the particle, respectively. 

\begin{figure*}[h!]
  \centering
  \includegraphics[width = 0.49\linewidth]{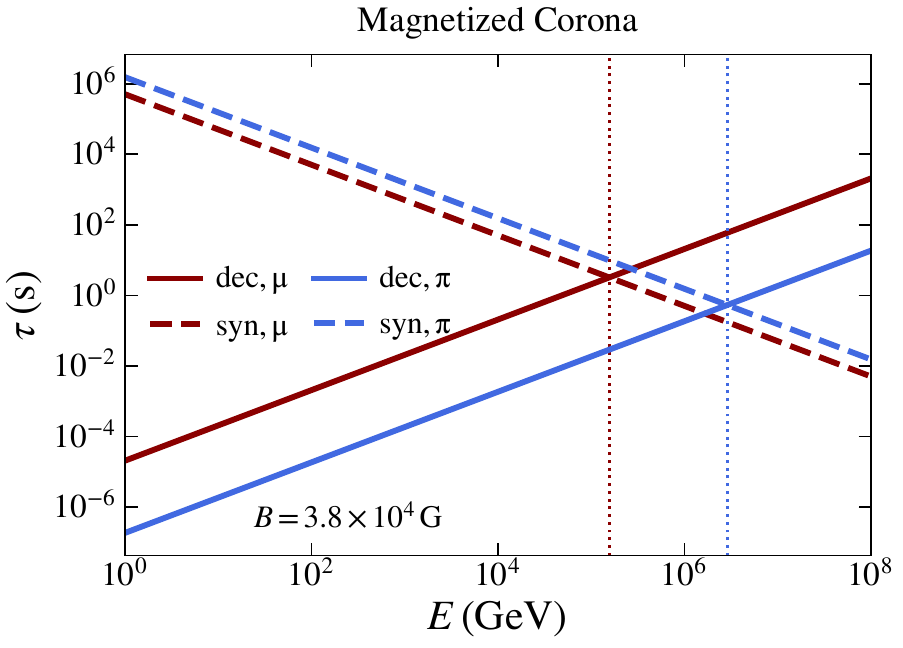}\includegraphics[width=0.49\linewidth]{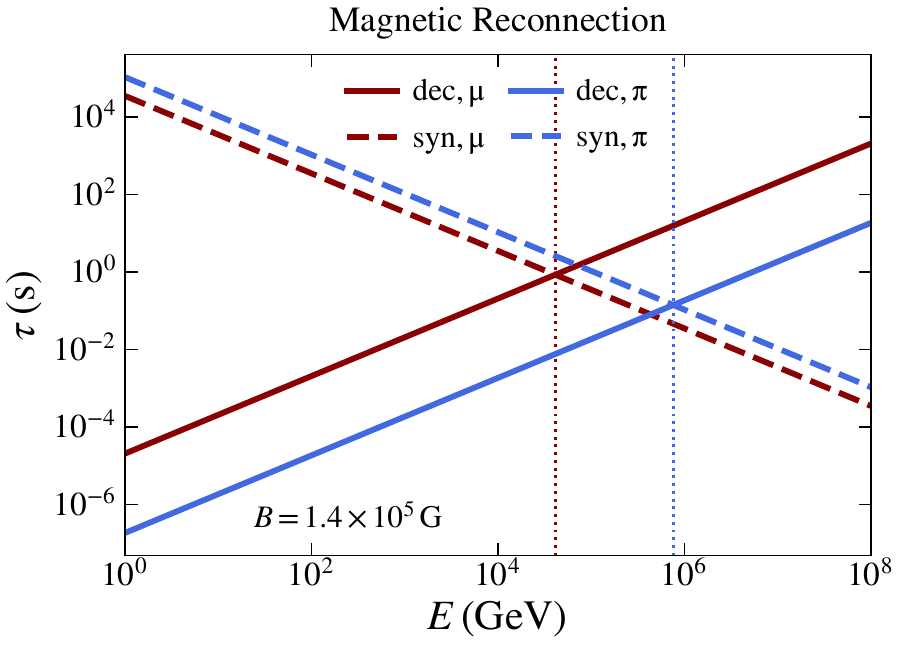}
  \caption{Decay (solid) and synchrotron-cooling (dashed) timescales of charged muons (dark red) and pions (blue) for the magnetized turbulent   corona (left) and magnetic-reconnection (right) configurations.  Vertical dotted lines mark the energies at which the two timescales  become equal for each particle species. The adopted magnetic fields are
  $B\simeq3.8\times10^4\,\mathrm{G}$ and
  $B\simeq1.4\times10^5\,\mathrm{G}$, respectively.}
  \label{fig:app_secondary_timescales}
\end{figure*}

As shown in Fig.~\ref{fig:app_secondary_timescales}, the critical muon energies are approximately $156\,\mathrm{TeV}$ for the magnetized turbulent corona and $41\,\mathrm{TeV}$ for magnetic reconnection. The corresponding values for charged pions are considerably higher, approximately $2.87\,\mathrm{PeV}$ and $0.76\,\mathrm{PeV}$, respectively. Since a neutrino from muon decay typically carries a fraction of order one third of its parent's energy, the characteristic neutrino energies associated with
muon cooling are $\sim 50\,\mathrm{TeV}$ and $\sim14\,\mathrm{TeV}$.

\subsection{Stationary transport of charged secondaries}
\label{app:secondary_transport}

In order to account for the possible energy losses of muons and pions, we solve their transport equation as

\begin{equation}
  -\frac{d}{dE}\!\left[b_{a,\mathrm{syn}}(E)N_a(E)\right]
  +\frac{N_a(E)}{\tau_{a,\mathrm{dec}}(E)}
  =Q_a^{\mathrm{inj}}(E),
  \label{eq:app_sec_transport}
\end{equation}
where $Q_a (E)$ denotes the injection rate. For pions, it comes directly from the photohadronic and hadronuclear collisions, while for muons it comes from the fraction of pions that decay rather than losing energy. They are specified in the next section. $N_a (E)$, instead, represents the equilibrium differential number density and it is given by 
\begin{equation} \label{eq:app_sec_transport_solution}
      N_a(E) = \frac{1}{b_{a,\mathrm{syn}}(E)}
  \int_E^\infty dE'\,Q_a^{\mathrm{inj}}(E')
  e^{-\mathcal{T}_a(E,E')},
\end{equation}
where 
\begin{equation}\label{eq:app_sec_transport_depth}
    \mathcal{T}_a(E,E')=\int_E^{E'}
  \frac{d\widetilde E}
  {b_{a,\mathrm{syn}}(\widetilde E)
   \tau_{a,\mathrm{dec}}(\widetilde E)}.
\end{equation}

\subsection{Injection of charged pions and muons}
\label{app:secondary_injection}
In order to calculate the injection of charged pions, we first note that neutrino emission is dominated by photohadronic interactions. Therefore, in what follows, we neglect the hadronuclear contribution. We adopt a channel-based response-function formalism
analogous to that implemented in the
AM$^3$ framework~\cite{Klinger:2023zzv}.
In particular, we employ the simplified photohadronic model \emph{Sim-B} of Ref.~\cite{Hummer:2010vx},
including resonance, direct, and multipion-production channels. Unlike AM$^3,$ which adopts the more detailed Sim-A prescription, our implementation uses Sim-B to provide a simplified treatment of secondary particle production and cooling. 
For either incident nucleon $N=p,n$, each interaction channel $i$ is
characterized by the average multiplicity
$M_{i,N}^{\pi^q}$ of a charged pion $\pi^q$ ($q=+,-$) and the fraction
$\chi_i$ of the incident nucleon energy carried by that pion.
The
charged-pion source term is
\begin{equation}\label{eq:app_sec_pion_source}
    Q_{\pi^q}^{\mathrm{inj}}(E_\pi)
  =\sum_{N=p,n}\sum_i\frac{M_{i,N}^{\pi^q}}{\chi_i}
  n_N\!\left(\frac{E_\pi}{\chi_i}\right)
  \times\Gamma_{i,N\gamma}
  \!\left(\frac{E_\pi}{\chi_i}\right),  
\end{equation}
where 
\begin{equation}\label{eq:app_sec_channel_rate}
     \Gamma_{i,N\gamma}(E_N)
  =\,c\int_0^\infty d\varepsilon\,n_\gamma(\varepsilon)
  f_i(\gamma_N\varepsilon).
\end{equation}
As in the main text, $\gamma_N=E_N/(m_Nc^2)$, while $n_N(E_N)$ represents the stationary nucleon density, and
$n_\gamma(\varepsilon)$ is the differential target-photon number density.
The so-called angle-averaged response function, $f$ reads
\begin{equation}
  f_i(y)=\frac{1}{2y^2}
  \int_{\varepsilon_{r,i}^{\mathrm{th}}}^{2y}
  d\varepsilon_r\,\varepsilon_r\sigma_i(\varepsilon_r),
  \label{eq:app_sec_simb_response}
\end{equation}
with $y=\gamma_N\varepsilon$. The parametrized cross sections, multiplicities, and energy fractions are
taken from the Sim-B channel tables of Ref.~\cite{Hummer:2010vx}.

Subsequently, the pions decay. Firstly, we determine the rate with which pions decay as
\begin{equation}
    Q_{\pi^{q}}^{\rm dec}(E_{\pi}) = \frac{N_{\pi^{q}}(E_\pi^{q})}{\tau_{\pi^{q}, \rm dec}(E_{\pi^{q}})}.
\end{equation}
For the muons, instead, the injection rate is 
\begin{equation}
  Q_{\mu_h^q}^{\mathrm{inj}}(E_\mu)
  =\int_{E_\mu}^\infty\frac{dE_\pi}{E_\pi}\,
  Q_{\pi^q}^{\mathrm{dec}}(E_\pi)
  \times F_{\pi^q\to\mu_h^q}
  \!\left(\frac{E_\mu}{E_\pi}\right),
  \label{eq:app_sec_muon_source}
\end{equation}
where $F_{\pi^q\to\mu_h^q}$ are helicity-dependent functions~\cite{Lipari:2007su} and $r_\pi\equiv(m_\mu/m_\pi)^2$. 
For these functions we follow Ref.~\cite{Lipari:2007su}
\begin{align}
  F_{\pi^+\to\mu_R^+}(x)
  &=F_{\pi^-\to\mu_L^-}(x)
  =\frac{r_\pi(1-x)}{(1-r_\pi)^2x},
  \label{eq:app_sec_pi_mu_R}\\
  F_{\pi^+\to\mu_L^+}(x)
  &=F_{\pi^-\to\mu_R^-}(x)
  =\frac{x-r_\pi}{(1-r_\pi)^2x},
  \label{eq:app_sec_pi_mu_L}
\end{align}
defined for $r_\pi \le x \le 1$ and vanish outside. Therefore, we solve the transport equations separately for right-handed and left-handed muons.

\subsection{All-flavor neutrino production}
\label{app:secondary_neutrinos}

Equipped with the pion and muons number densities, we determine the all-flavor neutrino flux following the approach of~\cite{Lipari:2007su}.
In particular, the neutrino production rate from the pions decay is given by

\begin{equation}  \label{eq:app_sec_neutrino_pion}
  Q_{\nu,\pi}(E_\nu)
  =\sum_{q=\pm}\int_{E_\nu}^\infty
   \frac{dE_\pi}{E_\pi}\,Q_{\pi^q}^{\mathrm{dec}}(E_\pi)
   \nonumber
  \times F_{\pi\to\nu}\!\left(\frac{E_\nu}{E_\pi}\right),
\end{equation}
where 
\begin{equation}
  F_{\pi\to\nu}(x)=\frac{1}{1-r_\pi},
  \qquad 0\leq x\leq1-r_\pi.
  \label{eq:app_sec_direct_pi_nu}
\end{equation}
The neutrino production rate from the muons decay, instead, is 
\begin{equation}\label{eq:app_sec_neutrino_muon}
  Q_{\nu,\mu}(E_\nu)
  =\sum_{q=\pm}\sum_{h=R,L}\int_{E_\nu}^\infty
   \frac{dE_\mu}{E_\mu}\,Q_{\mu_h^q}^{\mathrm{dec}}(E_\mu)
   \nonumber
  \times F_{\mu^q\to\nu,\mathrm{all}}
      \!\left(\frac{E_\nu}{E_\mu};h\right).
\end{equation}
The functions $F$ in this case are given by~\cite{Lipari:2007su}
\begin{equation}
  F_{\mu^+\to\nu,\mathrm{all}}(x;h)
  =\frac{11}{3}-9x^2+\frac{16}{3}x^3\nonumber + h\left(\frac{5}{3}-12x+21x^2-\frac{32}{3}x^3\right),
  \label{eq:app_sec_mu_all_kernel}
\end{equation}
 where $h=+1$ and $-1$ denote right- and
left-handed muons, respectively. For the charge-conjugate $\mu^-$ decay,
the sign of $h$ is reversed. The final all-flavor neutrino production rate is $Q_{\nu,\mathrm{all}}=Q_{\nu,\pi}+Q_{\nu,\mu}$.

\begin{figure*}[h!]
    \centering


    \begin{minipage}[t]{0.47\textwidth}
        \centering
        \includegraphics[width=\linewidth]{
            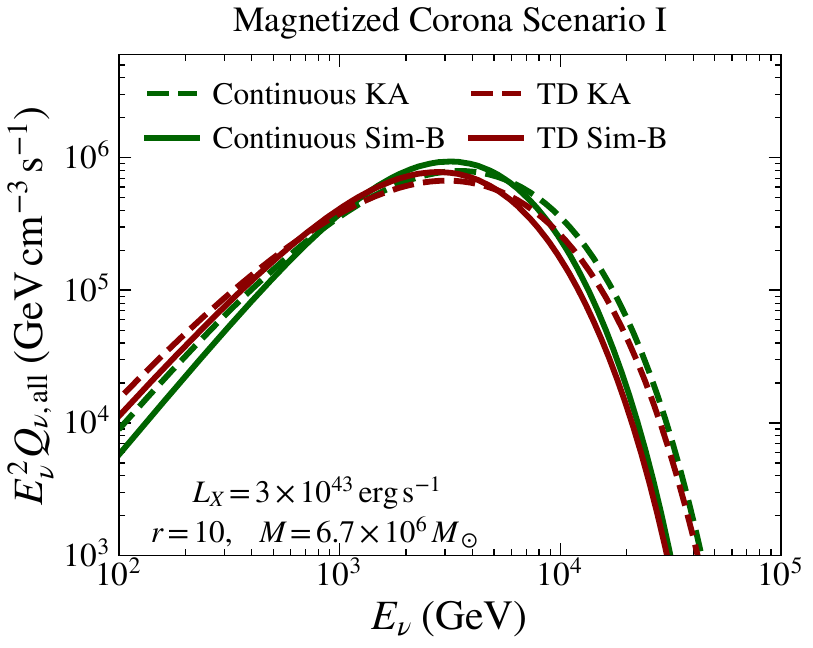
        }
    \end{minipage}
    \hfill
    \begin{minipage}[t]{0.47\textwidth}
        \centering
        \includegraphics[width=\linewidth]{
            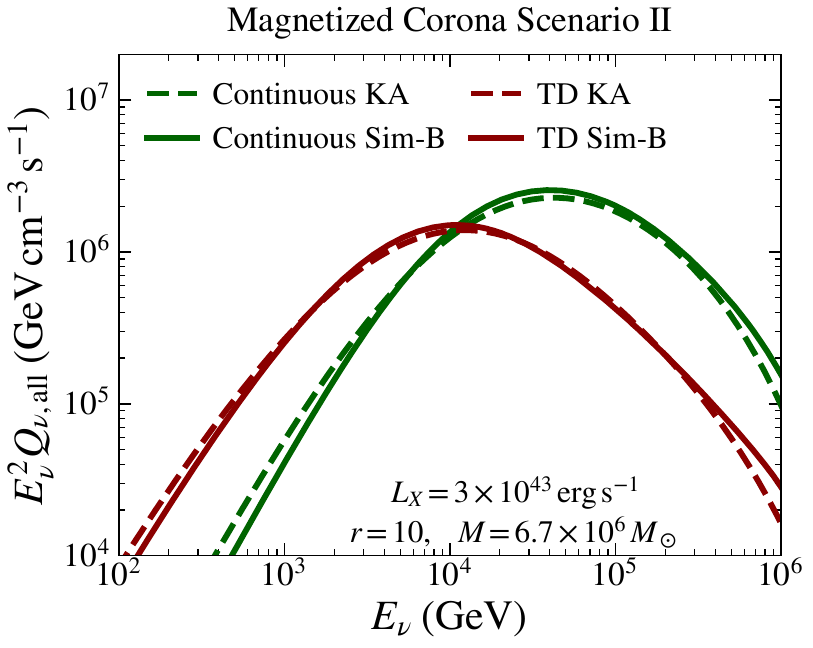
        }
    \end{minipage}

    \vspace{0.3cm}


    \begin{minipage}[t]{0.47\textwidth}
        \centering
        \includegraphics[width=\linewidth]{
            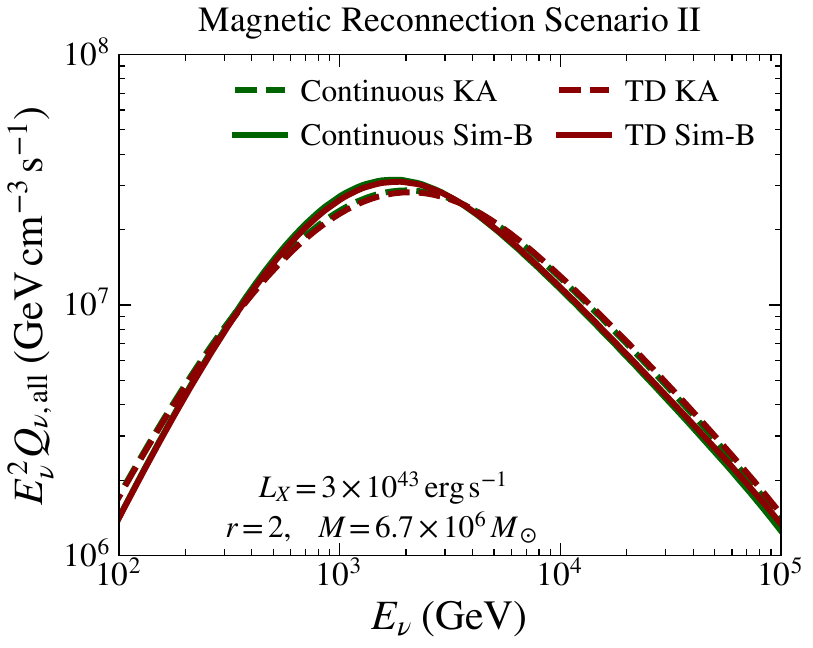
        }
    \end{minipage}

    \caption{
    Comparison of the all-flavor photohadronic neutrino
    production rates obtained using the Kelner--Aharonian
    (KA, dashed lines) and Sim-B (solid lines) parametrizations,
    neglecting synchrotron cooling of secondary particles.
    The upper panels correspond to the magnetized corona,
    considering scenarios I (left) and II (right), while the
    lower panel refers to the magnetic reconnection scenario.
    Dark-red and green lines represent the TD
    and CEL treatments, respectively.
    The overall agreement between the two parametrizations
    supports the use of the simplified Sim-B framework to
    investigate the impact of secondary cooling.
    }
    \label{fig:KA_SimB_comparison}

\end{figure*}

Figure~\ref{fig:KA_SimB_comparison} compares the all-flavor neutrino production rates obtained using the Kelner--Aharonian (KA) photohadronic parametrization~\cite{Kelner:2008ke}, adopted in the main text, with those computed using the simplified Sim-B approach introduced in this Appendix. For this comparison, synchrotron cooling of secondary pions and muons is neglected, such that $Q_{\pi}^{\rm dec}=Q_{\pi}^{\rm inj}$ and $Q_{\mu}^{\rm dec}=Q_{\mu}^{\rm inj}$. The two approaches show good agreement in both spectral shape and normalization, with differences in the integrated neutrino luminosities below approximately $10\%$ for the magnetized corona scenarios. This agreement supports the reliability of the simplified framework adopted here to investigate the impact of secondary cooling.

\begin{figure*}[h!]
    \centering

    \begin{minipage}[t]{0.47\textwidth}
        \centering
        \includegraphics[width=\linewidth]{
            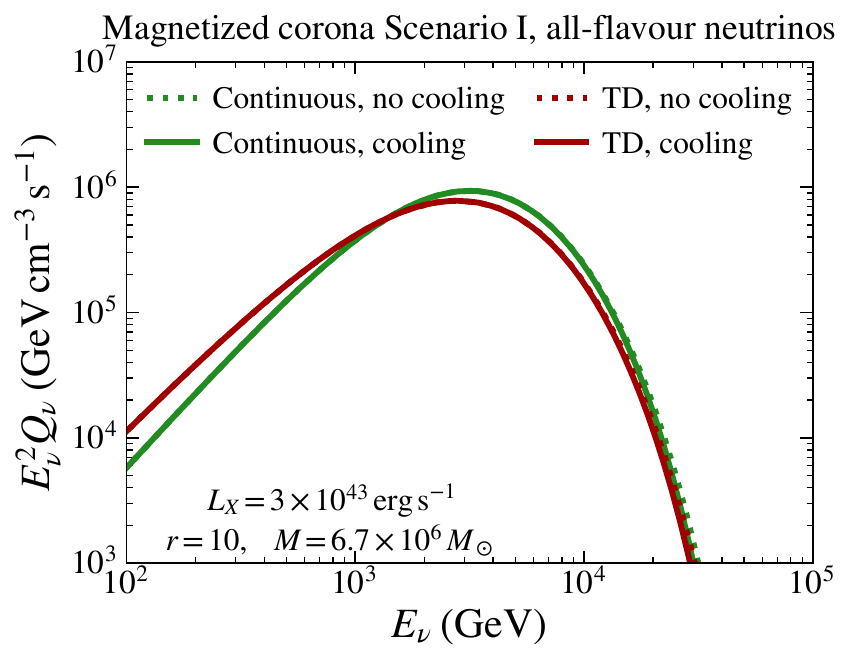
        }
    \end{minipage}
    \hfill
    \begin{minipage}[t]{0.47\textwidth}
        \centering
        \includegraphics[width=\linewidth]{
            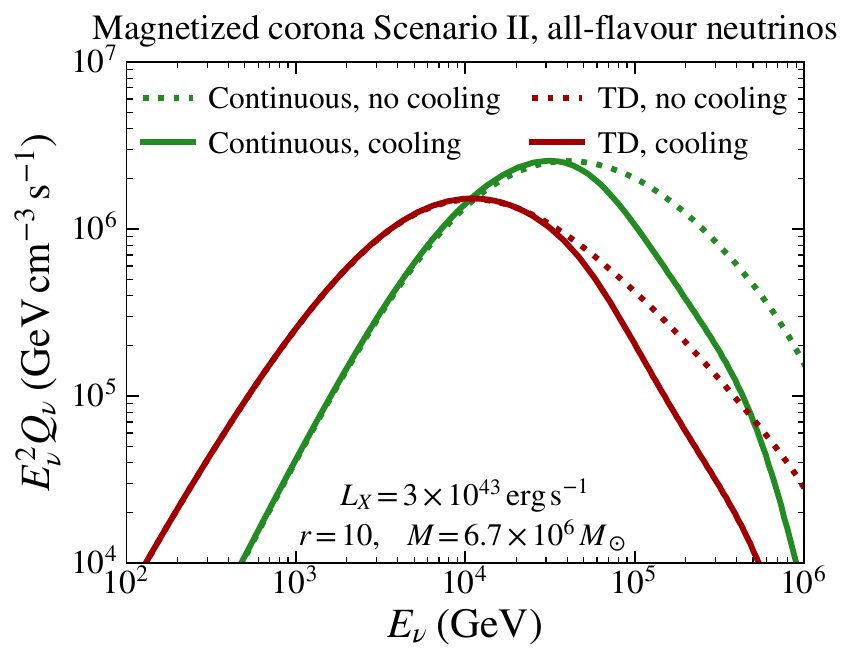
        }
    \end{minipage}

    \vspace{0.25cm}

    \begin{minipage}[t]{0.47\textwidth}
        \centering
        \includegraphics[width=\linewidth]{
            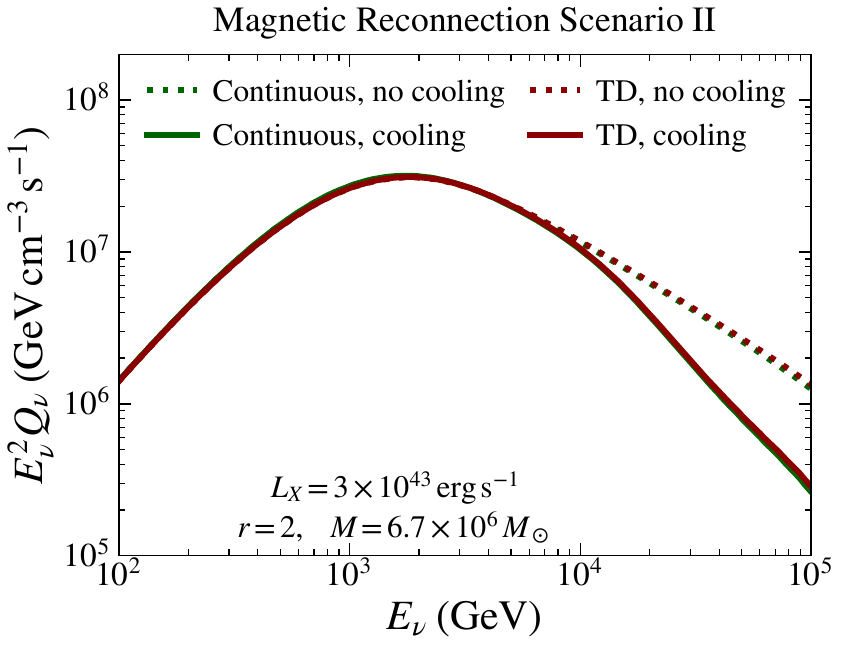
        }
    \end{minipage}

    \caption{
    All-flavor neutrino production rates obtained using
    the simplified Sim-B framework, with (solid lines)
    and without (dotted lines) synchrotron cooling of
    secondary charged pions and muons.
    The upper-left and upper-right panels correspond
    to scenarios I and II of the magnetized corona,
    respectively, while the lower panel refers to
    the magnetic reconnection scenario.
    Dark-red and green lines represent the 
    TD and CEL treatments, respectively.
    }
    \label{fig:secondary_cooling_comparison}

\end{figure*}

Fig.~\ref{fig:secondary_cooling_comparison} compares the all-flavor neutrino production rates obtained with (solid lines) and without (dotted lines) synchrotron cooling of secondary pions and muons. For Scenario I of the magnetized turbulent corona, the two treatments yield nearly identical neutrino spectra, indicating that secondary cooling has a negligible impact. For Scenario II, synchrotron cooling mainly suppresses the high-energy part of the neutrino spectra, with a more pronounced effect on the CEL solution. Nevertheless, the significant difference between the TD and CEL spectra persists. In the magnetic reconnection scenario, cooling primarily affects the neutrino emission above the spectral peak, leaving the TD and CEL solutions almost unchanged relative to each other.

Quantitatively, for Scenario II of the magnetized turbulent corona, the KA parametrization adopted in the main text predicts a reduction of 35\% in the integrated neutrino luminosity when comparing the TD and CEL treatments. Within the simplified Sim-B framework considered here, the corresponding reduction is approximately 39\% in the absence of secondary cooling, decreasing to about 30\% when synchrotron losses of pions and muons are included. The difference between the two predictions without cooling reflects the different photohadronic parametrizations. Therefore, we take the 30\% reduction as a representative estimate when secondary cooling is accounted for, confirming that the main conclusions remain unchanged. In the magnetic reconnection scenario, the corresponding difference remains at the percent level. Therefore, although secondary cooling modifies the neutrino spectra and their integrated luminosities, the main conclusions presented in the main text remain unchanged within the simplified framework considered here.

\clearpage

\bibliography{biblio}
\bibliographystyle{unsrt}

\end{document}